\documentclass[prd,preprint,aps,nofootinbib,superscriptaddress]{revtex4-1}

\usepackage{amsmath}
\usepackage{epsfig}
\usepackage{appendix}
\usepackage{subfigure}

\begin{document}


\title{Accidental Supersymmetric Dark Matter and Baryogenesis}

\author{Jonathan Kozaczuk}
\email{jkozaczu@ucsc.edu}\affiliation{Department of Physics, University of California, 1156 High St., Santa Cruz, CA 95064, USA}

\author{Stefano Profumo}
\email{profumo@ucsc.edu}\affiliation{Department of Physics, University of California, 1156 High St., Santa Cruz, CA 95064, USA}\affiliation{Santa Cruz Institute for Particle Physics, Santa Cruz, CA 95064, USA} 

\author{Carroll L. Wainwright}
\email{cwainwri@ucsc.edu} \affiliation{Department of Physics, University of California, 1156 High St., Santa Cruz, CA 95064, USA}

\date{\today}

\begin{abstract}
\noindent We show that ``accidental'' supersymmetry is a beyond-the-Standard Model framework that naturally accommodates a  thermal relic dark matter candidate and successful electroweak baryogenesis, including the needed strongly first-order character of the electroweak phase transition. We study the phenomenology of this setup from the standpoint of both dark matter and baryogenesis.  For energies around the electroweak phase transition temperature, the low-energy effective theory is similar to the MSSM with light super-partners of the third-generation quarks and of the Higgs and gauge bosons.  We calculate the dark matter relic abundance and the baryon asymmetry across the accidental supersymmetry parameter space, including resonant and non-resonant CP-violating sources. We find that there are regions of parameter space producing both the observed value of the baryon asymmetry and a dark matter candidate with the correct relic density and conforming to present-day constraints from dark matter searches. This scenario makes sharp predictions for the particle spectrum, predicting a lightest neutralino mass between 200 and 500 GeV, with all charginos and neutralinos within less than a factor 2 of the lightest neutralino mass and the heavy Higgs sector within 20-25\% of that mass, making it an interesting target for collider searches. In addition, we demonstrate that successful accidental supersymmetric dark matter and baryogenesis will be conclusively tested with improvements smaller than one order of magnitude to the current performance of electron electric dipole moment searches and of direct dark matter searches, as well as with IceCube plus Deep Core neutrino telescope data.
\end{abstract}

\maketitle
\section{Introduction}

The Standard Model (SM) of particle physics is missing several key ingredients needed for a satisfactory phenomenological description of nature.  First, it does not provide an explanation for the observed baryon asymmetry of the universe (BAU).  
Second, the SM does not contain any viable particle candidates for dark matter (DM), which is needed to explain a large array of astrophysical and cosmological observations.  From a more theoretical perspective, the SM additionally falls short of explaining the large hierarchies between fundamental physical scales.  In particular, it provides no satisfactory explanation for why the Planck scale, $M_{\rm Pl}\sim 10^{19}$ GeV, is so much higher than the electroweak (EW) scale, $m_{\rm EW}\sim 100$ GeV.  This is known as the hierarchy problem.   

In recent years, several models have been suggested that address the hierarchy problem.  In particular, models with warped extra dimensions, such as Randall-Sundrum (RS) scenarios, have been proposed which naturally generate the hierarchy between the Planck and EW scales \cite{Randall:1999ee}.  In RS models, the universe is described by a five-dimensional (5D) geometry with two four-dimensional (4D) branes located at the UV (Planck) and IR (TeV) points.  The Higgs fields are localized on (or near) the IR brane, and the warped fifth dimension ``redshifts'' the Planck scale to the TeV scale, providing a rather elegant solution to the hierarchy problem.  Additionally, by placing the SM fermions in the bulk, hierarchies between the Yukawa couplings can be accounted for by the wave-function overlap with the Higgs boson in the fifth dimension \cite{Gherghetta:2000qt}.  

While explaining the hierarchy problem, simply embedding the SM in a RS scenario is not fully satisfactory.  To prevent sizable CP-violating effects from Kaluza-Klein (KK) modes in the absence of additional flavor structure, the IR scale must be at or above $\mathcal{O}(10$ TeV$)$\cite{KK_modes}.  Precision electroweak experiments also dictate that the IR scale must be larger than the EW scale, hence some additional tuning is required between these scales.  This is an incarnation of the so-called \emph{little} hierarchy problem.  To resolve this issue, models of ``emergent" or ``accidental" supersymmetry have been proposed (see e.g. Refs.~\cite{Gherghetta:2003wm, Sundrum:2009gv, accidental}), in which supersymmetry (SUSY) emerges as an accidental symmetry in the IR, with SUSY broken on the UV brane.  As a result, the Higgs mass can be protected from radiative corrections up to the IR scale, while the warped extra dimension generates the hierarchy between the TeV and Planck scales.  Within this framework, which we describe in more detail in Sec.~\ref{sec:Accidental_SUSY}, both hierarchy problems can potentially be resolved.  The specific particle content of the theory depends on the model of SUSY embedded in the Randall-Sundrum spacetime.  Since we are interested in the general features of accidental supersymmetric models, we will consider the particle content of the minimal supersymmetric extension of the standard model (MSSM) as a conservative case from the standpoint of the field content of the theory.  
 
Randall-Sundrum scenarios can do more than just solve the big and little hierarchy problems.  In fact, we show here that models with warped extra dimensions may also provide an explanation for the origin of the baryon asymmetry via the mechanism of electroweak baryogenesis (EWB). Crucial to this mechanism are a number of conditions, ultimately related to those generically needed for any dynamical mechanism for the production of a baryon asymmetry \cite{Sakharov:1967dj}: first, one needs departure from thermal equilibrium at the electroweak scale; second, one needs large enough charge (C) and charge-parity (CP) violation; third, one needs violation of baryon number. The second and third conditions are easily satisfied in the context of supersymmetric EWB models: any minimal supersymmetric extension to the SM, in fact, contains numerous (albeit constrained) new sources of CP violation, while baryon number (B) violation, is provided by SM weak sphalerons --- we will comment on this more below. More critical is how to have a large deviation from thermal equilibrium --- a condition in practice realized, in the context of EWB, via a strongly first-order electroweak phase transition. As pointed out in Ref.~\cite{Nardini:2007me}, here the RS setup may be of crucial importance, as we also explain below.

While the universe is described by the Randall-Sundrum spacetime at zero temperature, at finite temperature, RS models possess an additional high-temperature phase, described by an Anti-de Sitter-Schwarzschild (AdS-S) spacetime with a black hole horizon replacing the TeV brane \cite{Creminelli:2001th}.  Alternatively, a holographic description facilitated by the AdS-CFT correspondence also exists in which the two phases correspond to a deconfined and to a confined phase of a strongly coupled gauge theory, respectively.  Provided that the free energy of the RS phase is less than the free energy of the AdS-S phase, $F_{RS}<F_{AdS-S}$, a phase transition can occur between the two: as the universe cools below a temperature $T_c$, bubbles of the TeV brane can begin to nucleate out of the black hole horizon \cite{Creminelli:2001th} (see also Refs.~\cite{Randall:2006py, Kaplan:2006yi, Nardini:2007me} for further discussion of the confining phase transition).

Because, from the CFT perspective, conformal invariance is only spontaneously broken in the RS phase, $F_{RS}>F_{AdS-S}$ implies that the RS phase is metastable.  For the confining phase transition to occur, one must introduce some mechanism to explicitly break conformal invariance.  From the AdS perspective, this can be accomplished by stabilizing the radion (the field governing the separation between the UV and IR branes) with a potential generated e.g. by additional 5D fields.  Once the radion is stabilized, the free energy of the two phases will be equal at some temperature $T=T_c$ producing a phase transition via bubble nucleation with nucleation temperature $T_n\leq T_c$.  In many cases, $T_n$ can be significantly lower than the temperature of the electroweak phase transition (EWPT) predicted by the 4D Minkowski theory \cite{Nardini:2007me}.  Since the Higgs sector is typically confined to the IR brane, this low nucleation temperature results in a ``{\em supercooled}'' EWPT (i.e. taking place at lower temperatures than otherwise possible), thereby potentially strengthening the phase transition.  While this supercooling was studied specifically in the case of the SM embedded in RS with a Goldberger-Wise potential \cite{Goldberger:1999uk} for the radion in Ref.~\cite{Nardini:2007me}, this possibility is a consequence of the geometry and localization of the Higgs sector in the IR and is largely independent of the particle content of the theory and can therefore potentially arise in accidental SUSY as well.  As a result, models of accidental SUSY may provide a strongly first order EWPT even \emph{without} e.g. a light right-handed scalar top (stop) quark \cite{Carena:2008vj, Carena:2008rt}, or additional singlets contributing to the Higgs potential  \cite{Profumo:2007wc}, as is typically required for successful EWB in the MSSM.  Alternatively, as we explain in the next section, certain incarnations of the accidental SUSY framework also posit, as a solution to the $\mu$-problem, an additional extension to the Higgs sector via a singlet scalar field. This potentially provides an additional route to a strongly first order electroweak phase transition.

Since a strongly first order phase transition appears to be a natural possibility in accidental SUSY, the remaining issue pertinent to EWB is the requirement of large enough CP-violation to seed weak sphalerons, which will be the primary focus for the rest of this study.  In fact, accidental SUSY naturally satisfies this requirement as well.  Even in its MSSM incarnation, there are several new CP-violating phases which can source the baryon asymmetry. In particular, there are new phases arising in the higgsino-gaugino and third-generation scalar sectors.  It has recently been shown that of the third-generation scalars, only CP-violating stau sources can account for the observed baryon asymmetry while still conforming to various phenomenological and experimental constraints from electric dipole moment searches \cite{Kozaczuk:2012xv}.  Here we concern ourselves with moderate values of the ratio of Higgs vevs, $\tan\beta =10$, in which case the stau sources are suppressed.  We will therefore be interested in EWB with higgsino-gaugino sources in accidental SUSY.  Electroweak baryogenesis utilizing these sources in the MSSM has been extensively analyzed in recent studies (e.g. \cite{Huet:1995sh, Carena:1996wj, Lee:2004we, Chung:2008aya, Lepton_Mediated, Supergauge, Including_Yukawa, Konstandin:2003dx, Konstandin:2004gy, More_Relaxed, Carena:2002ss, Carena:2008vj, EWB_and_EDMs, Balazs:2004ae, EWB_and_DM, Kozaczuk:2011vr, Menon:2004wv, Huber:2006wf}), and we build on these analyses in our study of the accidental SUSY scenario.    Note that extending the particle content beyond that of the MSSM would provide more potential  sources of CP-violation.

Supersymmetric RS models also have the added benefit of generically containing a viable dark matter candidate, if the lightest supersymmetric particle (LSP) corresponds to the lightest neutralino, over some regions of parameter space.  This is a result of $R$-parity conservation, whereby the LSP is stable.  Thus it may be possible for accidental SUSY to simultaneously explain the origin of the BAU and the nature of dark matter, while also solving both the big and little hierarchy problems.  In fact, the production of both the relic DM density and the baryon asymmetry via higgsino-gaugino sources are closely connected \cite{EWB_and_DM, Kozaczuk:2011vr}, since both depend predominantly on the higgsino mass term $\mu$ and on the gaugino soft-supersymmetry breaking masses $M_1$ and $M_2$.  Consequently, enforcing both the correct DM properties and baryon asymmetry in conformity with various observational constraints may result in sharp predictions for the regions of interest within the accidental SUSY parameter space.  This is an attractive possibility and one which we explore in the present study.  

In what follows we compute the baryon asymmetry across the parameter space of a minimal (MSSM-like) incarnation of accidental SUSY.  We do so independently of the specifics of SUSY breaking, choosing higgsino and gaugino masses which yield the correct DM relic density (this is the so-called ``well-tempered neutralino" setup \cite{welltempered}) and assuming a strongly first-order electroweak phase transition arising either from the supercooling provided by the AdS-S transition or from the contribution of a gauge singlet super-field to the effective potential.  We then impose constraints from electric dipole moment (EDM) measurements and from dark matter searches to outline potentially viable regions of the parameter space.  We also consider the impact of the projected sensitivities of these various experiments.  In doing so, we find that accidental SUSY models, even in their most minimal incarnations, may allow for successful EWB and a viable DM candidate provided that the resulting soft breaking wino mass, $M_2$, and the higgsino mass parameter $\mu$ are roughly degenerate, with the LSP a bino-higgsino admixture; we also require the MSSM heavy Higgs sector to lie relatively close to twice the LSP mass.  These findings can be used to hone in on specific models of accidental SUSY giving rise to the observed properties of our universe. 

\section{Accidental Supersymmetry}\label{sec:Accidental_SUSY}

The framework of accidental SUSY can be used to both generate a hierarchy between the Planck and IR scales and a little hierarchy between the electroweak and IR scales by embedding supersymmetry in an RS spacetime \cite{accidental}.  The Higgs mass is protected from higher order corrections up to the IR scale $m_{\rm IR}$ by requiring the superpartners of the third-generation quarks\footnote{Typically, the RH sbottom is taken to be heavy, as sbottom loop contributions to the Higgs mass are subdominant}, as well as those of the gauge and Higgs bosons, to be light in order to cancel the loop corrections from their SM counterparts to the Higgs mass.  The warped extra dimension then takes over in protecting the hierarchy between the IR and Planck scales for scales above $m_{\rm IR}$.   On the other hand, superpartners of the quarks and leptons of the first two generations should be heavy to avoid excessive flavor and CP-violation which would occur if the mediation of SUSY breaking is not flavor blind.  This spectrum is reminiscent of so-called ``Split SUSY" models \cite{split_susy}.  

Much previous work has been devoted to the study of supersymmetric Randall-Sundrum models \cite{Gherghetta:2000qt, Casas:2001xv, Gherghetta:2003wm,Goh:2003yr, Sundrum:2009gv, accidental}; the specifics of the resulting particle spectrum inherently rely on the underlying assumptions about particle content, SUSY breaking, localization of the particles in the 5D spacetime, etc.  However, here we are concerned with the generic features of accidental supersymmetry relevant to EWB and to dark matter phenomenology, and so we limit our assumptions to a few key points representing the main features of this setup.  For the sake of generality, we take the particle content embedded in the RS spacetime to be the minimal supersymmetric spectrum of the MSSM.  Our assumptions about the resulting spectrum, typical of accidental SUSY models, are listed in the bullet points below.  For the sake of illustration, we focus in this section on the model set forth in Ref.~\cite{accidental} as a concrete example demonstrating how these features arise in accidental SUSY scenarios, briefly and qualitatively summarizing some of the aspects of the setup relevant for our investigation below.  We stress, however, that the remainder of this study will not necessarily depend on this particular model.  We work in the 5D gravity language of the gauge-gravity duality except where noted.  Readers comfortable with the general setup of accidental SUSY may skip down to our bullet points below.  

We consider a minimal supersymmetric theory embedded in a RS spacetime with metric \begin{equation} ds^2=e^{-2k\left|y\right|}\eta_{\mu \nu} dx^{\mu} dx^{\nu} + dy^2 \end{equation} where $\eta_{\mu\nu}$ is the Minkowski metric and $k$ denotes the scale of AdS curvature.  The 5th dimension is an $S^1/Z_2$ orbifold with coordinate $y$.  The points $y=0,\ell$ are the positions of the UV and IR branes, respectively, corresponding to the orbifold fixed points.  Denoting the 4- and 5-dimensional Planck masses as $M_4$, $M_5$, respectively, the warped-down IR scale on the $y=\ell$ brane is assumed to be $m_{\rm IR}=e^{-k\ell}k$ (we assume $k\ell\sim 30$ throughout), and we assume the UV cutoff on the IR brane is given by $\Lambda_{\rm IR}=e^{-k\ell}M_5$.  In Ref.~\cite{accidental}, to naturally implement the split SUSY spectrum, supersymmetry is broken on the UV brane at an intermediate scale $M_{\rm SUSY} \ll M_5$, with light standard model fermions localized near the UV brane so that their superpartners feel SUSY-breaking maximally.  This results in heavy first- and second-generation sfermions.  Meanwhile, the higgsinos and stops are localized near the IR brane so that they remain light.  The gauginos are protected by an accidental $R$-symmetry (they have sizable wavefunction overlap with the UV brane, and so would typically be heavy without this symmetry).  One can introduce both (i) a bulk hypermultiplet, which obtains an $F$-term when SUSY breaking occurs from its coupling to a SUSY-breaking spurion on the UV brane, and (ii) a constant superpotential on the IR brane\footnote{A constant superpotential must also be added to the UV brane to tune the cosmological constant to zero.} with mass scale $C$.  Combined, both (i) and (ii) generate a potential for the radion (whose chiral superfield is denoted by $\omega$), stabilizing the radion at a scale $m_{\rm IR}$ which we take to be around $10$ TeV  (see Ref.~\cite{accidental} for details concerning the radion potential in this scenario).  

We are ultimately interested in the low energy phenomenology of accidental SUSY models, below the cutoff $\Lambda_{\rm IR}$.  To extract the low energy effective theory, one must integrate over the extra dimension and account for the effects of the radion superfield by canonically normalizing the radion and chiral superfields localized on the IR brane.  This yields masses for the scalar (and pseudoscalar) degrees of freedom (associated with the real and imaginary parts of the complex radion field, respectively) near the soft scale at the minimum of the potential \cite{accidental}.  Denoting this canonically normalized radion  superfield as $\varphi$,  one can show \cite{Casas:2001xv}, given certain ans\"{a}tze, that the effective low energy superpotential on the IR brane, $W^{\rm IR}_{\rm eff}$, reads, after canonical normalization, \begin{equation} \label{eq:superpotential} W^{\rm IR}_{\rm eff} = e^{-3\varphi/\Lambda_{\rm IR}}W^{\rm IR}(Q_i, H_i), \end{equation} where $W^{\rm IR}$ is the superpotential for the IR-localized chiral superfields (the Higgs and third-generation quarks in our case).  As a result, any dimensionful coupling in the superpotential of the embedded supersymmetric theory, which is naturally near the 4D Planck scale, will be warped down to the IR scale.  In the case of an MSSM-like superpotential $W^{\rm IR}_{\rm MSSM}$, employed here, this results in $\mu\sim m_{\rm IR}$ and thus a supersymmetric little hierarchy must be reintroduced so that $\mu$ is near the soft scale, as required to obtain the correct dark matter relic abundance (discussed in Sec.~\ref{sec:welltempered}).  We comment more on this below, but for the moment we shall simply assume some tuning so that $\mu\sim100-1000$ GeV.

The effective superpotential $W^{\rm IR}_{\rm eff}$ in Eq.~\ref{eq:superpotential} shows that there can be considerable differences between the low energy phenomenology of a supersymmetric RS theory and its 4D flat-space counterpart due to the presence of the radion superfield.  The K\"{a}hler potential is also typically non-minimal in such theories \cite{Casas:2001xv}.  The effective superpotential and K\"{a}hler potential can result in mixing between the radion and Higgs, between the radion superpartner (the ``radino") and higssinos, as well as couplings of the radion fields to the other IR-localized degrees of freedom.  While these effects are interesting from the standpoint of low-energy phenomenology, they are largely model-dependent and so we neglect them for the remainder of our study.  In particular, we assume that: (1) the radino mass (which depends on the details of SUSY breaking) is large enough so that it decouples from the phenomenology, and (2) that the IR-localized fields couple minimally to gravity so that there is no mixing between the Higgs and radion.  While these assumptions still allow for couplings of the radion to the IR-localized fields (and thus potentially impacting the calculation of the baryon asymmetry), we estimate that the relevant rates are small compared to other processes of interest (see Eq.~\ref{eq:radion_rate} and the surrounding discussion).  In addition to the radion, RS models are accompanied by the usual Kaluza-Klein excitations, which can also play a role in the phenomenology.  However, as discussed above, we assume that these modes are heavy, in order to avoid constraints from e.g. CP- and flavor-violation; this assumption effectively decouples them from the processes of interest here.  Summarizing, as a result of these assumptions, \emph{the particle content we consider is simply that of the MSSM, with the mass hierarchy typical of accidental SUSY models}, discussed below.  One should bear in mind that effects associated with the radion which we neglect here can significantly impact other aspects of low-energy phenomenology, such as those associated with collider searches \cite{radion_collider}.      

The main virtue of accidental SUSY models for our purposes, then, is that they can provide an attractive supersymmetric spectrum for the sectors relevant for dark matter and electroweak baryogenesis, in addition to facilitating the requirement of a strongly first-order EWPT.  With SUSY breaking occurring on the UV brane, the various parts of the visible sector feel SUSY breaking in different ways.  In Ref.~\cite{accidental}, SUSY breaking for the UV-localized supermultiplets is felt primarily from the effects of heavy UV-localized messengers.  Integrating out the messengers (with mass scale $M_X$), yields the 5D effective Lagrangian \cite{accidental} \begin{equation} \mathcal{L}_5 \supset \delta(y)\int d^4\theta \frac{\Phi^{\dagger}\Phi}{k^2 M_X^2}Q^{\dagger}Q, \end{equation} where $\Phi$ is the SUSY-breaking spurion.  Upon inserting the wavefunction of the zero-mode for the multiplet $Q$, one obtains \begin{equation} \label{eq:m_uv} m^{\rm UV}_{\rm soft}\sim \frac{M^2_{\rm SUSY}}{M_X} \end{equation} for the soft masses of UV-localized sfermions.  For reasonable choices of the messenger scale, these large masses effectively decouple them from the processes of interest for electroweak baryogenesis, which occur below the TeV scale. 

In order to protect the Higgs mass from gauge boson loops, the gauginos must also be light, implying that the generically leading contribution to the gaugino masses $\Phi W_{\alpha}W^{\alpha}$ must be suppressed; otherwise, the gauginos would acquire soft masses $m_{\rm gaugino}$ of the same order as Eq.~\ref{eq:m_uv}.  This can be done, as in Ref.~\cite{accidental}, by charging the spurion $\Phi$ under a $U(1)'$ gauge symmetry, in which case the leading contribution to $m_{\rm gaugino}$ typically arises from \begin{equation} \mathcal{L}_5 \supset \delta(y)\int d^4 \theta \frac{\Phi^{\dagger} \Phi}{k^2 M_X^3}W^{\alpha}W_{\alpha} + {\rm h.c.} .\end{equation}  Inserting the zero-mode gaugino wavefunction yields gaugino soft masses \begin{equation} \label{eq:m_gaugino} m_{\rm gaugino}\sim \left(\frac{M^2_{\rm SUSY}}{M_X^2 k\ell} \right) m^{\rm UV}_{\rm soft} \end{equation} which can be of order the soft IR scale $m^{\rm IR}_{\rm soft}$ provided that $M_X\sim M_{\rm SUSY}^{4/3}/m^{1/3}_{\rm IR}$.  The SUSY-breaking scale is confined to be in the range $10^7$ GeV $\lesssim M_{\rm SUSY} \lesssim 10^{11}$ GeV to obtain a sufficiently large $m^{\rm UV}_{\rm soft}$ while preventing excessive contributions from radion mediation to the gaugino masses \cite{accidental}.

Finally, the IR-localized supermultiplets must also be light to stabilize the electroweak scale.  If these particles are sufficiently localized on the IR brane, the largest contributions to the corresponding soft masses arise from gravity mediation.  These contributions arise from the part of the 4D effective Lagrangian given by \cite{accidental} \begin{equation} \mathcal{L}_4\supset \int d^4\theta \omega^{\dagger}\omega \frac{\left[\Phi^{\dagger}\Phi\right]_{\rm IR}}{M_5^3}\left(Q^{\dagger}Q +H^{\dagger}_u H_u+H^{\dagger}_d H_d\right) \end{equation} where we have considered the Higgs sector of the MSSM (this differs from Ref.~\cite{accidental} as already mentioned).  The above Lagrangian leads to IR soft masses \begin{equation} \label{eq:m_IR} m^{\rm IR}_{\rm soft} \sim \sqrt{6} \left(\frac{C}{M_5}\right)^3 m _{\rm IR}\end{equation} which can provide the little hierarchy between soft and IR scales for only a modest hierarchy between the scale of the constant IR superpotential and the 5D Planck scale. 

As is well known, in the MSSM, some tension exists between having light stops and obtaining a Higgs mass consistent with the (tentative) 125 GeV Higgs reported by CMS and ATLAS \cite{higgsclaims}.  A way out is having significant mixing in the stop sector, which can allow for a heavy enough Higgs without decoupling the stops (see e.g. Ref.~\cite{Djouadi:2005gj}).  This possibility would not change the phenomenology relevant for calculating the baryon asymmetry or dark matter density (although it would require some additional tuning).  On a related note, in considering the particle content of the MSSM embedded in the RS spacetime, we have not ameliorated the so-called ``$\mu$-problem", which reintroduces a supersymmetric little hierarchy as mentioned above, since $\mu$ will typically be of the order of the warped-down Planck scale, $m_{\rm IR}$ which is necessarily higher than the EW scale.  This could be addressed by considering e.g. the NMSSM Higgs sector (as in Ref.~\cite{accidental}), which dynamically gives rise to $\mu$ near the soft IR scale, while at the same time raising the tree-level Higgs mass so that tuning of the stop mixing is not required.  In fact, the additional gauge singlet superfield is not crucial to the phenomenology we are interested in here and so we could very well frame our discussion in the NMSSM (provided that the LSP is not singlino-like).  However, we content ourselves with considering the particle content of the MSSM despite these issues, specifically to emphasize that one does not necessarily require the singlet introduced in the NMSSM for successful baryogenesis in accidental supersymmetric models, as discussed in the introduction.  This fact, along with the minimal set of additional CP-violating phases in the MSSM, suggests that MSSM-like accidental SUSY scenarios provide a conservative look at the prospects for accidental supersymmetric dark matter and baryogenesis.

We note that in addition to the soft-breaking masses in Eqs.~\ref{eq:m_uv}, \ref{eq:m_gaugino}, \ref{eq:m_IR}, there are also hard-SUSY breaking terms in the effective 4D Lagrangian which affect the couplings of the gauginos (such as those governing CP-violating higgsino-gaugino source and the so-called ``supergauge interactions" discussed in Sec.~\ref{sec:BAU}) and (scalar)$^4$ interactions \cite{Sundrum:2009gv}.  These enhanced couplings will affect the light scalar masses through loop corrections, and a mild tuning must generally be invoked to keep these effects small, implying corrections $\lesssim \mathcal{O}(10\%)$ to the couplings \cite{accidental}.   While we neglect these (model-dependent) corrections in our calculations, the reader should bear in mind that larger gaugino couplings will enhance the baryon asymmetry, strengthen the EDM constraints, and increase the various dark matter cross-sections, thereby potentially strengthening the exclusions discussed in Sec.~\ref{sec:paramspace} for models with large hard-breaking effects. 

The above considerations are only one specific realization of the accidental supersymmetric framework.  However, some general features emerge for the spectrum typical of such theories.  Specifically, accidental SUSY naturally accommodates: \begin{itemize} 
\item Light third generation squarks, gauginos, and higgsinos, $m^{\rm IR}_{\rm soft}$, $m_{\rm gaugino}\lesssim 1$ TeV;
\item Heavy sleptons and first- and second-generation squarks ($m^{\rm UV}_{\rm soft}\gtrsim 1000$ TeV);
\item Heavy Kaluza-Klein modes, possibly starting at around $40$ TeV;
\item A radion at the soft scale 
\item A neutralino LSP 
\item Model-dependent masses for the heavy Higgs sector, gravitino, and radino. 
\end{itemize}  
The effective 4D theory at low energies we will consider is described by the MSSM with the above spectral features (we assume that the effects of the gravitino, radion fields and KK modes on processes of interest are negligible).  Rather than considering one detailed model of SUSY breaking and the resulting mass patterns, we proceed model-independently in choosing values for the various relevant parameters in our calculations as described in the following sections.  As we will see, the above features of the typical accidental SUSY spectrum are attractive from both the standpoint of dark matter and electroweak baryogenesis.

\section{A Well-Tempered Neutralino}\label{sec:welltempered}

In the context of minimal supersymmetric extensions to the Standard Model with heavy sfermion masses and a neutralino LSP, as in the low-energy effective theories of accidental SUSY of interest here, the thermal relic density of the LSP is fixed by (i) the relevant entries in the neutralino mass matrix: the higgsino mass term $\mu$,  the bino soft supersymmetry breaking mass term $M_1$, and the wino soft supersymmetry breaking mass term $M_2$; and (ii) the presence or absence of a resonant annihilation channel via the light ($h$) or heavy $(H,\ A)$ Higgses, or with the $Z$ boson. The accidental SUSY scenario does not imply either a rigid hierarchy among $\mu,\ M_1$ and $M_2$, or a specific mass range for the lightest neutralino or for the heavy Higgs sector. We therefore take here the model-independent view of treating all the relevant parameters as free, while at the same time enforcing the requirement of a thermal relic density matching the cold dark matter density, $\Omega_\chi h^2\simeq \Omega_{\rm DM}\simeq 0.11$. 

Numerous studies have addressed the set of $(\mu,\ M_1,\ M_2)$ producing a ``well-tempered'' thermal relic neutralino \cite{welltempered}: early analyses of bino-higgsino mixing generating the right thermal relic density include e.g. Refs.~\cite{ref22, ref23, ref16, ref24, ref25}, while wino-bino mixing was originally studied, to our knowledge, in Refs.~\cite{ref25, ref26, ref27}. With heavy sfermion masses, bino-like neutralinos have very suppressed pair-annihilation cross section, making it indispensable to have either some degree of higgsino- or wino- mixing, a resonant annihilation channel, or one (or more) co-annihilation partner(s). Generically, mixed higgsino-wino neutralinos have masses well above one TeV, with lighter, sub-TeV higgsino-wino neutralino LSPs being systematically under-abundant as thermal relic dark matter candidates. Such heavy LSPs push the mass of the particles relevant to the CP violating sources responsible for EWB to exceedingly large values, making them too heavy to produce the observed baryon asymmetry.

In this study, we explore the $(\mu,\ M_1,\ M_2)$ parameter space by focusing on the $(M_1,\ M_2)$ plane, where we calculate (using the DarkSUSY code \cite{darksusy} for the computation of the neutralino thermal relic density) the value of $\mu$ that leads to the correct thermal relic density. Since, as we will show, low (meaning at or below a TeV) values for the mass-scale of the heavy Higgs sector will be generically needed to produce a large enough BAU, we choose for the sake of illustration the two values $m_A=500$ GeV and $m_A=1000$ TeV (note that significantly smaller values of $m_A$ lead to excessive SUSY contributions to e.g. $b\to s\gamma$, while larger values fail to yield a large enough BAU, as we show below). This choice will lead to resonances for neutralino masses $m_\chi\simeq m_A/2\simeq 250$, 500 GeV.

\begin{figure*}[!t]
\mbox{\hspace*{-1.2cm}\includegraphics[width=0.55\textwidth,clip]{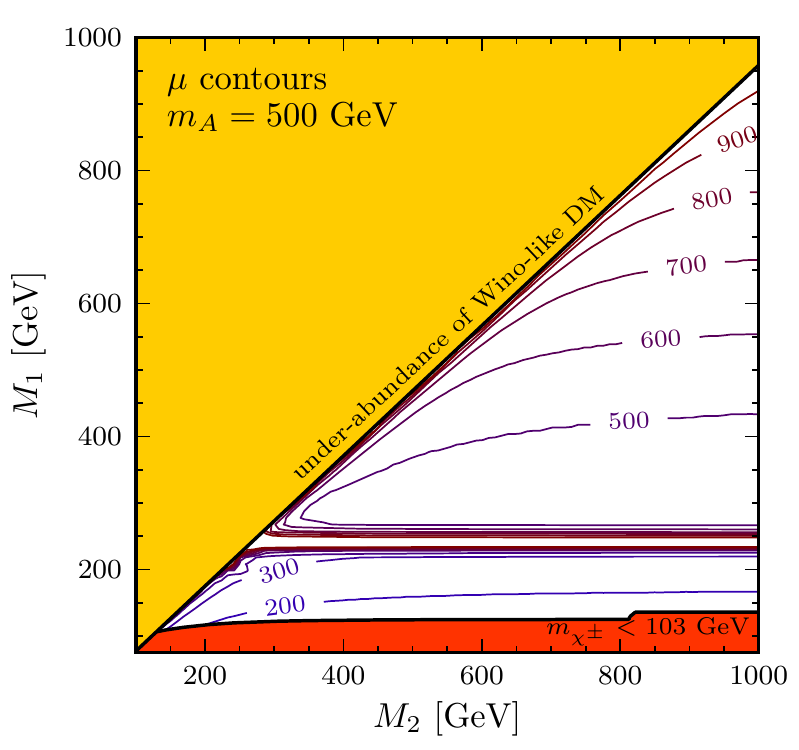}\quad\includegraphics[width=0.55\textwidth,clip]{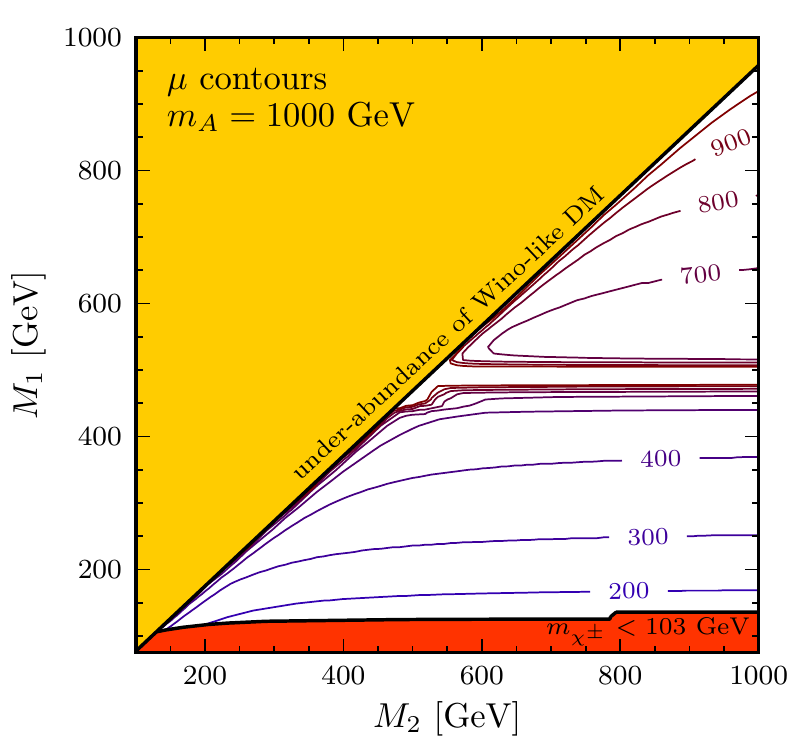}}
\caption{\label{fig:isomu}\it\small Curves of constant $\mu$ (in GeV) on the ($M_1,M_2$) plane corresponding to ``well-tempered'' neutralinos producing a thermal relic density matching the observed cold dark matter density. The red regions are ruled out by LEP searches for charginos, while the yellow region has wino-like neutralinos with a largely under-abundant density for any value of $\mu$. In the left panel we set $m_A=500$ GeV, while in the right panel to 1000 GeV.}
\end{figure*}

Fig.~\ref{fig:isomu} presents the results of the procedure outlined above. The yellow region with $M_2<M_1$ features wino-like neutralinos, with $\Omega_\chi h^2\ll\Omega_{\rm DM}h^2$ for any value of $\mu$. The red region at the bottom of the plot has charginos lighter than the LEP limit of about 103 GeV, and is thus ruled out (note that at present LHC searches do not significantly constrain this parameter space in a generic way, i.e.\ not assuming any relation between gaugino masses, and thus for generic gluino masses). The resonances appear for $M_1\simeq m_\chi\simeq m_A/2$. Additionally, large values of $\mu$ correspond to the $M_1\simeq M_2$ region, where bino-wino mixing efficiently suppresses the abundance of relic neutralinos. The generic feature of the plots is that for each value of $M_1$ a value of $\mu\gtrsim M_1$ is selected, with larger $\mu$ for smaller $M_2$, where bino-wino mixing starts being important.

In the remainder of this study, we employ the values of $\mu$ shown in Fig.~\ref{fig:isomu}, thus enforcing the correct thermal relic density, and vary quantities (such as the CP violating phases) that affect very marginally the thermal relic density (see e.g. the discussion in Ref.~\cite{EWB_and_DM}) while being crucial to the BAU calculation. We then compute the BAU produced by EWB (Sec.~\ref{sec:BAU}), electric dipole moments (Sec.~\ref{sec:EDM}), and dark matter detection rates (Sec.~\ref{sec:DM}), comparing model predictions with existing experimental data and with the performance of future experiments.

\section{The Baryon Asymmetry}\label{sec:BAU}
In supersymmetric models of electroweak baryogenesis, the baryon asymmetry is produced by baryon number-violating weak sphalerons, acting on a net left-handed (LH) fermionic charge density $n_L$.  The LH density arises from CP-violating interactions of the particles with the EWPT bubble wall.  For a detailed and recent review of EWB in supersymmetry, see Ref.~\cite{Morrissey:2012db}.

At the practical level, the calculation of the baryon asymmetry in supersymmetric EWB requires two steps: first, a set of coupled quantum Boltzmann equations (QBEs) must be solved for the production and transport of the chiral charge; second, since weak sphalerons are essentially decoupled from the system of QBEs, to obtain the baryon asymmetry $Y_{\rm B}$ one simply integrates over the LH charge density in the unbroken phase, where weak sphalerons are active\footnote{Weak sphalerons are inactive in the broken phase precisely because of the required strongly first order phase transition}.  While a detailed study of the electroweak phase transition in accidental SUSY would be interesting and necessary to establish the details surrounding e.g. the nucleation temperature in these models, it is beyond the scope of this paper and so we simply assume the necessary strongly first order electroweak phase transition in our calculations as a result of either the supercooling by the AdS phase transition or of some additional particle content not related to the mechanism of EWB (e.g. an additional singlet superfield in the superpotential). 

To solve the set of QBEs for  $n_L$, one must keep track of all relevant particle number-changing interactions active near the EWPT.  These consist of: \begin{enumerate} \item CP-violating sources.  The relevant CP-violating phases for the low-energy accidental SUSY effective theory near the EWPT are those of the MSSM in our setup.  As discussed in the Introduction, the sources we consider are those associated with higgsino-gaugino-vev interactions.  \item Tri-scalar and Yukawa interactions between the quark and Higgs superfields.  These rates arise from the interactions in the MSSM superpotential, as well as through the soft breaking effects discussed in Sec.~\ref{sec:Accidental_SUSY}.  \item Chiral relaxation rates for the (s)quark and higgsino densities.  These rates are CP-conserving, arising from the interaction of the various fields with the Higgs vevs at the EWPT bubble wall.  They tend to dilute the overall baryon asymmetry. \item Supergauge rates, responsible for converting between the SM particles and their corresponding superpartners.  If these rates are fast, then chemical equilibrium is established between SM densities and their superpartner counterparts.  Throughout our calculations, we follow previous studies \cite{ Kozaczuk:2011vr, Lee:2004we, Chung:2008aya, Lepton_Mediated} and assume superequilibrium for all relevant particle species in the transport equations, so that we can consider common densities for the interacting particles and their superpartners. \item Strong sphalerons, which convert third generation quarks to the first- and second generation, and vice versa.  Strong sphalerons can efficiently erase the net chiral charge generated by the CP-violating sources if the stop masses are heavy \cite{Giudice:1993bb, Moore:1997im}.  As a result, we assume that the stops are light enough so that this suppression does not occur, but also heavy enough not to be the LSP.  This is in fact a natural feature of accidental SUSY scenarios, and one of its virtues from the standpoint of EWB.  For our numerical calculations of the baryon asymmetry, as in previous studies \cite{Kozaczuk:2011vr}, we assume a RH stop soft mass $m_{U_3}=0$ GeV and LH soft mass $m_{Q_3}=1000$ GeV to show the maximal extent of the parameter space compatible with the observed baryon asymmetry in this setup.  
\item Interactions involving the radion.  As discussed in Sec.~\ref{sec:Accidental_SUSY}, these interactions should be generically small relative to the usual MSSM-like rates, since for particles of mass $m$ they are typically suppressed by powers of $m/m_{\rm IR}$ (or $T/m_{\rm IR}$ at high temperatures).  For example, the radion couples to the trace of the energy momentum tensor of the IR-localized fields \cite{Goldberger:1999un, Csaki:1999mp}, which results in Higgs-like interactions proportional to the masses of the various particles but suppressed by $m_{\rm IR}$.  A simple estimate comparing the rates for radion-top and top Yukawa interactions at finite temperature $T$, yields \begin{equation} \label{eq:radion_rate}    \frac{\Gamma_{\varphi t}}{\Gamma_{yt}} \simeq \left(\frac{T}{y_t m_{\rm IR}}\right)^2 \lesssim 10^{-4} \end{equation}  where $y_t$ is the top Yukawa coupling and we have assumed $T$ is much larger than the top thermal mass.  Eq.~\ref{eq:radion_rate} implies that the radion interaction rates are typically small, and so we neglect them in our calculation of the baryon asymmetry. One should bear in mind that any particle-changing relaxation rate, such as those involving the radion, will tend to suppress the overall baryon asymmetry as $1/\sqrt{\Gamma}$.  \end{enumerate}   With the above considerations, \emph{the calculation of the baryon asymmetry in our effective accidental SUSY theory is analogous to the calculation in the MSSM with light third generation squarks, higgsinos, and gauginos}, and the relevant interaction rates then are simply those of the MSSM.  A more detailed account of the rates in 1-5 above is provided in Ref.~\cite{Supergauge}, to which we refer the interested reader.

With the above considerations, we calculate the baryon asymmetry following the techniques and assumptions detailed in Refs.~\cite{Lee:2004we, Including_Yukawa, Chung:2008aya, Lepton_Mediated, Supergauge}.  The evaluation of the CP-violating sources, as well as the CP-conserving chiral relaxation rates, is carried out using the so-called Higgs ``vev-insertion approximation", in which interactions between the particles and the spacetime-varying Higgs vevs in the bubble wall are treated perturbatively \cite{Carena:1997gx}.  This prescription leads to a resonance in both the CP-violating and conserving rates for roughly degenerate particle masses for the relevant species involved.  Schematically, the higgsino gaugino source is given by terms of the form \begin{equation} \label{eq:source} \begin{aligned} S^{\rm CPV}_{\tilde{H}i}&= \frac{g_i^2}{\pi^2}v(x)^2\dot{\beta}(x) \operatorname{Arg}(M_i \mu)\\ &\times \int_0^{\infty}\frac{dk k^2}{\omega_{\tilde{H}}\omega_i}\operatorname{Im}\left\{ \frac{n_F(\mathcal{E}_i)-n_F(\mathcal{E}^*_{\tilde{H}})}{(\mathcal{E}_i-\mathcal{E}^*_{\tilde{H}})^2}-\frac{n_F(\mathcal{E}_i)+n_F(\mathcal{E}_{\tilde{H}})}{(\mathcal{E}_i+\mathcal{E}_{\tilde{H}})^2}\right\} \end{aligned} \end{equation} where $\omega^2_{\tilde{H},i}\equiv \left| \bf{k}\right|^2+M_{\tilde{H},i}$, $\mathcal{E}_{\tilde{H},i}\equiv \omega_{\tilde{H},i} - i\Gamma_{\tilde{H},i}$ (here the $\Gamma_{\tilde{H},i}$ are the thermal widths of the corresponding particles in the plasma), $n_F$ is the Fermi distribution function, and the index $i$ denotes the various quantities for the wino or bino contributions to either the neutral or charged sources.  From the structure of Eq.~\ref{eq:source}, we see that in our setup the CP-violating sources are strongest in parameter space regions where either the bino or wino soft mass ($M_1$ or $M_2$) is nearly degenerate with the higgsino mass parameter $\mu$ so that the corresponding denominator $\mathcal{E}_i-\mathcal{E}^*_{\tilde{H}}$ in Eq.~\ref{eq:source} is small, and where the relevant particles are light (c.f. the Boltzmann suppression factors in Eq.~\ref{eq:source}). 

A few comments on these calculational  techniques are in order.  First, the vev-insertion approximation tends to overestimate the production of the overall baryon asymmetry: in considering an approximate all-orders re-summation of the Higgs vev-insertions in perturbation theory, Refs.~\cite{Carena:2000id,Carena:2002ss} showed that the resonance exhibited in Eq.~\ref{eq:source} is smoothed out (hence suppressed) by the resummation.  Second, the resummation techniques of Refs.~\cite{Carena:2000id,Carena:2002ss} show that there are other, non-resonant contributions to $S^{\rm CPV}$ not appearing to lowest order in the vev-insertion approximation.  These contributions are the dominant ones away from the resonance and for large values of the pseudoscalar Higgs mass, $m_A$, which we discuss below.  The drawback of these approximately resummed sources is that it is not clear whether or not they are consistent with the power-counting done to calculate the relevant relaxation rates in the closed-time-path formalism \cite{Lee:2004we} (for more details concerning the different existing techniques for evaluating the baryon asymmetry in the literature, see Ref.~\cite{Morrissey:2012db}).  Since these non-resonant sources may open up more parameter space for EWB, we consider their impact separately in Sec.~\ref{sec:nonressrc}: we show there that our conclusions obtained in the vev-insertion approximation are qualitatively largely unchanged.

From the above form for the higgsino-gaugino source, Eq.~\ref{eq:source}, we see that accidental SUSY, with its prediction of light higssinos and gauginos, can result in a sizable baryon asymmetry through the source terms described above, provided the relevant mass terms are nearly degenerate.  Additionally, as mentioned above, the prediction of a rather light RH stop in accidental SUSY (c.f. Eq.~\ref{eq:m_IR}), also fares well for EWB, since a light stop is required to prevent the efficient erasure of chiral charges by strong sphaleron processes\footnote{Note that the suppression of the BAU with heavy stops can be ameliorated by considering so-called ``lepton-mediated" scenarios of EWB \cite{Lepton_Mediated}}.  These considerations, in conjunction with the generic strongly first-order phase transition potentially provided by the RS geometry, suggest that successful electroweak baryogenesis may be naturally and successfully accomplished in accidental SUSY models.  The question remains: what regions of the accidental SUSY parameter space are most likely to produce the observed baryon asymmetry while satisfying available experimental and observational constraints?

\begin{figure*}[!t]
\mbox{\hspace*{-1.2cm}\includegraphics[width=0.55\textwidth,clip]{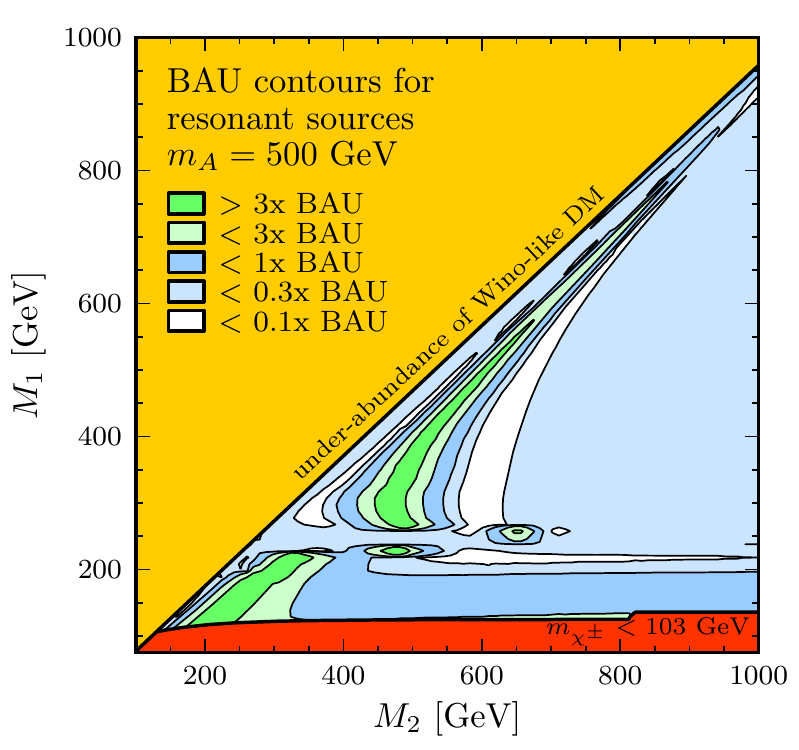}\quad\includegraphics[width=0.55\textwidth,clip]{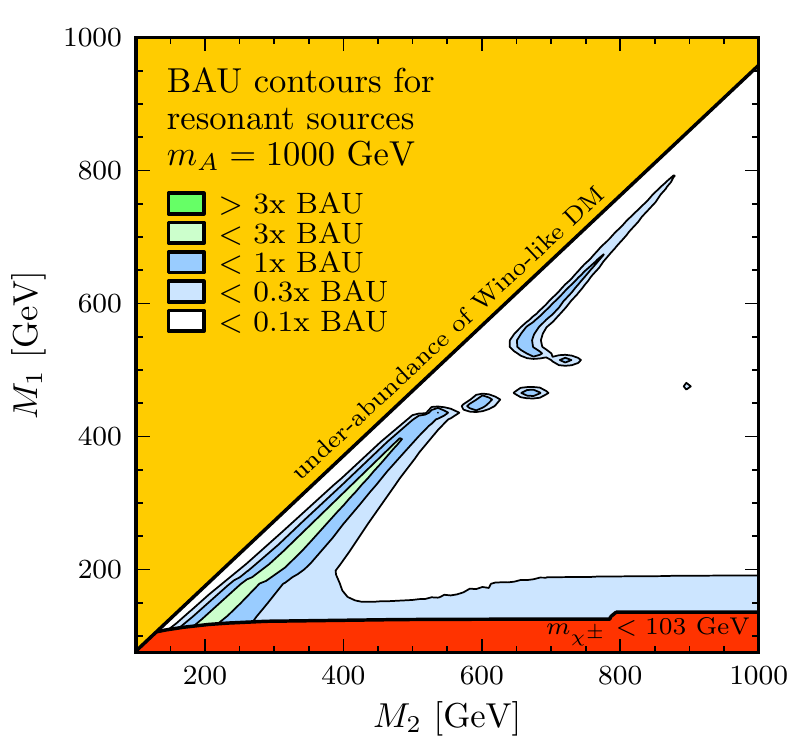}}
\caption{\label{fig:EWB}\it\small Curves of constant BAU (in units of the observed BAU) for the ``well-tempered'' neutralinos of Fig.~\ref{fig:isomu} for $m_A=500$ GeV (left) and 1000 GeV (right), assuming maximal CP-violating phases and the values of the various other parameters discussed in the text.}
\end{figure*}

To address this question, we calculate the baryon asymmetry produced by higgsino-gaugino CP-violating interactions with the EWPT bubble wall.  We vary the masses of the gauginos, $M_{1,2}$, while fixing $\mu$ by requiring the correct DM relic density, as discussed in Sec.~\ref{sec:welltempered}.  We show the resulting regions compatible with the observed baryon asymmetry ($Y_{\rm Obs}$) for $m_A=500, 1000$ GeV, maximal CP-violating phases, and with the choices for the other parameters discussed above in Fig.~\ref{fig:EWB} on the left and right, respectively.  Comparing Figs.~\ref{fig:isomu} and \ref{fig:EWB}, we see that the asymmetry is largest near the wino-higgsino resonance as expected, with a bino-like LSP.  We discuss the resulting implications for viable accidental SUSY EWB and DM in light of the other relevant DM and EDM constraints in Sec.~\ref{sec:paramspace}. 

There are several theoretical uncertainties associated with the production of the baryon asymmetry at the electroweak phase transition.  Dependence on the bubble wall parameters (the velocity, thickness, and variation of the ratio of Higgs vevs, $\Delta \beta$, across the wall) can introduce $\mathcal{O}(10-100$ GeV) uncertainties in the constant-BAU contours in the gaugino mass planes.  For some of these parameters, the effect on $Y_B$ is simple - for example, $Y_B$ is linear in $\Delta\beta$ for the resonant sources used here.  For the values of $m_A$ we consider, $\Delta \beta$ ranges from $(0.2- 2) \times 10^{-3}$ (using the two-loop results of Ref.~\cite{Moreno:1998bq}).  The dependence on the wall width and velocity is not as straightforward, since these quantities enter into other terms of the transport equations besides the CP-violating source, as well as in the integral over $n_L$.  The BAU generally decreases with increasing $L_w$ and is maximized for values of $v_w$ around a few $\times 10^{-2}$ (large velocities render the transport of chiral current inefficient, while smaller velocities lead to a quasi-equilibrium situation, also suppressing the asymmetry) \cite{ Huber:2001xf,Kozaczuk:2011vr}.  In our numerical calculation of the BAU we choose the optimistic values $v_w=0.05$, $L_w = 5/T$ to estimate the maximal extent of the EWB-compatible parameter space.  

An additional, significant, but to our knowledge less appreciated, uncertainty on $Y_B$ in the present calculational framework is that associated with the nucleation temperature, $T_n$, around which the processes relevant for EWB occur.  In Fig.~\ref{fig:EWB}, we assumed $T_n=100$ GeV; however, the nucleation temperature can in principle be lower or higher than this value, and without a more detailed and model-dependent study of the EWPT in accidental SUSY models, its value is at best known to an order of magnitude.  

\begin{figure}[!t]
\mbox{\hspace*{1.5cm} \includegraphics[width=1\textwidth,clip]{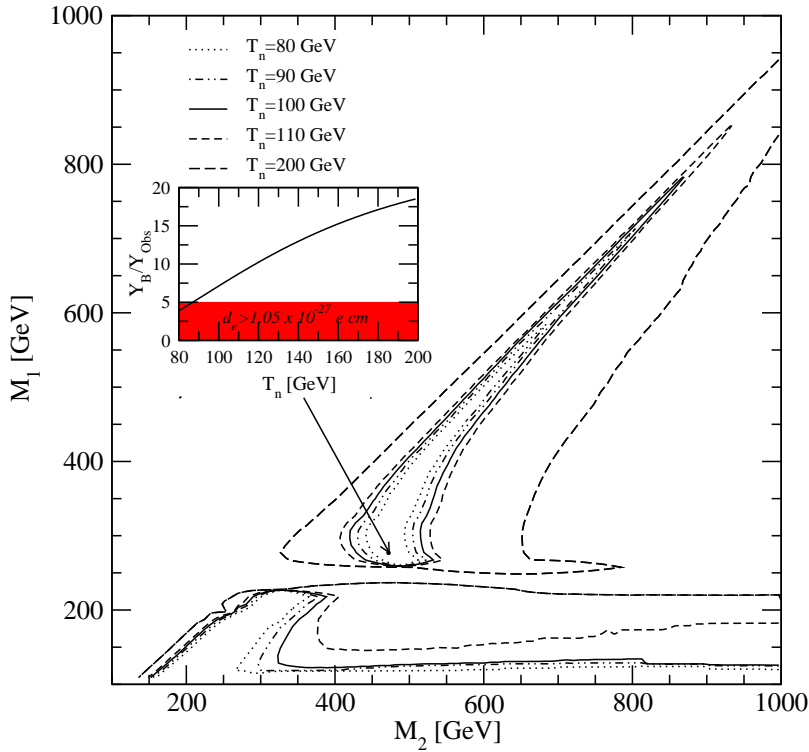}}
\vspace*{-1.5cm}
\caption{\label{fig:temp_dep}\it\small Curves of $Y_B=Y_{\rm Obs}$ for different nucleation temperatures and for $m_A=500$ GeV with maximal CP-violating phase.  The inset shows the temperature dependence of the resulting baryon asymmetry for a point near the resonance; the red shaded region is excluded by the non-observation of the electron EDM (discussed in Sec.~\ref{sec:EDM}) .  Smaller nucleation temperatures reduce the weak sphaleron rate and result in larger Boltzmann suppression while larger temperatures can enhance the BAU.}
\end{figure}

It is however possible to quantitatively assess the impact of this uncertainty: we compute the baryon asymmetry for different values of $T_n$ in Fig.~\ref{fig:temp_dep}.  Lower temperatures reduce the baryon asymmetry, as the sphaleron rates are ``slower'' and the Boltzmann suppression stronger in this regime.  Conversely, larger temperatures enhance the BAU.  Fig.~\ref{fig:temp_dep} implies that if the EWPT is made strongly first order by the mechanism of supercooling described above, the resulting nucleation temperature must not be too low ($\gtrsim 80$ GeV), otherwise much of the potentially viable parameter space for EWB and DM discussed in Sec.~\ref{sec:paramspace} will be ruled out\footnote{This can be viewed as an upper limit on the number of inflationary e-folds surrounding the phase transition as discussed e.g. in Ref.~\cite{Nardini:2007me}}.  We encourage the reader to bear this caveat in mind in interpreting our results in the following sections.   

\section{Electric Dipole Moments}\label{sec:EDM}

The general MSSM contains 40 CP-violating phases in addition to the single CP-violating phase in the standard model CKM matrix. These generally give rise to EDMs and chromo-EDMs in elementary fermions, nucleons, and neutral atoms. The current non-observation of any such EDMs puts stringent constraints on beyond-the-standard-model physics (for a recent study of constraints on CP-violating phases from EDM searches see e.g. Ref.~\cite{Li:2010ax}).

In our model, all relevant one-loop single-particle EDMs are suppressed by the large masses of the first and second-generation sfermions. Since we consider only CP-violation in the phases of $M_1$ and $M_2$ (or, equivalently, in $\mu$)\footnote{Technically, the physical CP-violating phases correspond here to $\phi_{M_{1,2}}\equiv{\rm Arg}(\mu M_{1,2}b^*)$, with $b$ the soft SUSY-breaking Higgs mass parameter.}, there are no contributions to chromo-EDMs. Instead, the dominant contributions come from two-loop Barr--Zee-like diagrams~\cite{Barr:1990vd} involving chargino-neutralino loops. The electron-EDM provides the most stringent constraint on our model, with an experimental bound of $|d_e| < 1.05\times10^{-27} e$-cm (coming from experiments on the YbF molecule)~\cite{Hudson:2011zz}. The current constraint from the neutron-EDM is also quite strong ($|d_n| < 2.9\times10^{-26} e$-cm)~\cite{Baker:2006ts}, but tends to be about 30\% weaker than the electron constraint relative to our model's predictions. On-going experiments may improve the sensitivity to the electron EDM by up to two orders of magnitude (see, e.g., Ref.~\cite{Hewett:2012ns}), which has the potential to constrain almost the entire parameter space for baryogenesis in an accidental SUSY model. A non-null observation at that sensitivity level could also point to new physics consistent with this model.

\begin{figure*}[!t]
\mbox{\hspace*{-1.2cm}\includegraphics[width=0.55\textwidth,clip]{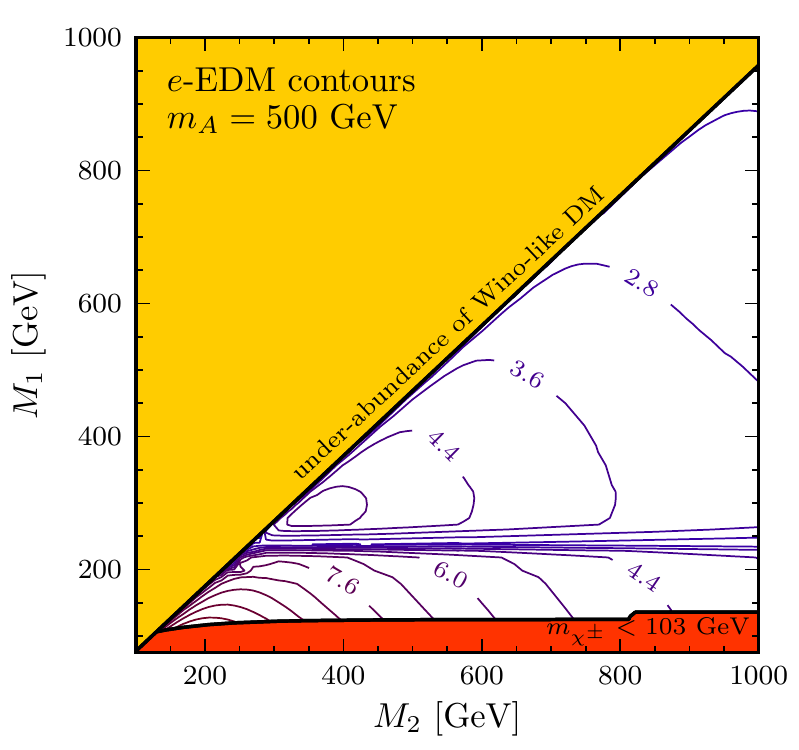}\quad\includegraphics[width=0.55\textwidth,clip]{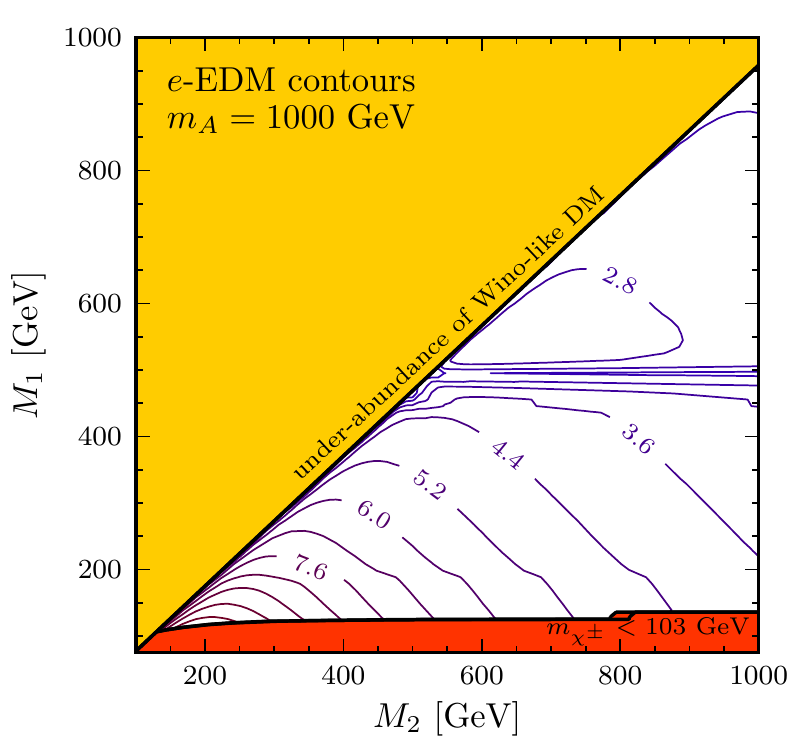}}
\caption{\label{fig:EDM}\it\small 
Curves of constant electron-EDM for $m_A=500$ GeV (left) and 1000 GeV (right) and maximal CP-violating phase, $\phi_{M_1}=\phi_{M_2} = \pi/2$. The labels are in terms of the current experimental bounds: $d_e/d_{e-bound}$, where $d_{e-bound} = 1.05\times10^{-27} e$-cm. Curves of constant neutron-EDM (not shown) are qualitatively similar, but provide less stringent constraints.}
\end{figure*}

We use the expressions in Ref.~\cite{Li:2008kz} to calculate the electron and neutron EDMs in our model, along with the FeynHiggs package~\cite{FeynHiggs} to calculate the Higgs mass and mixing angles including the full effects of CP violating phases. Fig.~\ref{fig:EDM} shows curves of constant electron-EDM along the $M_1$--$M_2$ plane with maximal CP-violating phases, $\phi_{M_1}=\phi_{M_2} = \pi/2$. At each point, the value of $\mu$ is taken from Fig.~\ref{fig:isomu} to provide the correct dark matter relic abundance. The experimental bounds on both electron and neutron-EDMs rule out the entire plotted parameter space {\em for maximally CP-violating phases}. Of course, smaller CP-phases are viable: the appropriate size of the CP violating phase depends on the requirement of matching the observed BAU, as calculated, for  $\phi_{M_1}=\phi_{M_2} = \pi/2$, in Fig.~\ref{fig:EWB}. We postpone the calculation of the resulting EDM constraints to our summary section on the accidental SUSY parameter space in Sec.~\ref{sec:paramspace}.

\section{Direct and Indirect Dark Matter Searches}\label{sec:DM}

The phenomenology of neutralino dark matter in the incarnation of the MSSM corresponding to the accidental SUSY framework described above depends, generically, on a relatively small set of parameters. These include the relevant mass scales entering the neutralino mass matrix ($\mu,\ M_1,\ M_2$) and the mass scale of the heavy Higgs sector (e.g. fixed by the physical mass $m_A$). Other light particles, including the radion, radino, and stops, are largely non-influential, as long as none of those particles is the LSP. Our analysis of the ($M_1,M_2$) plane therefore satisfactorily exhausts the relevant dark matter phenomenology for the model under study.

\begin{figure*}[!t]
\mbox{\hspace*{-1.2cm}\includegraphics[width=0.5\textwidth,clip]{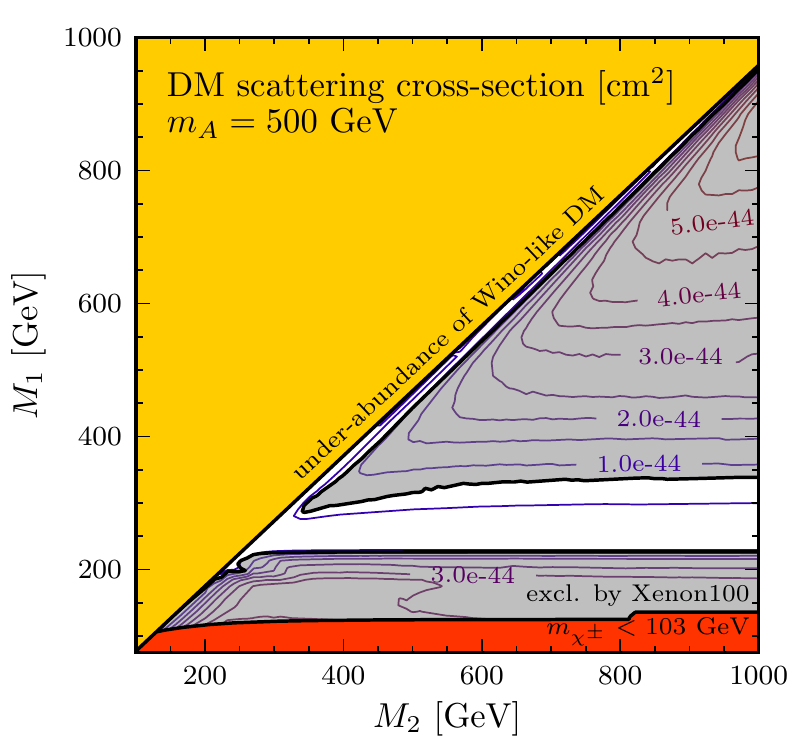}\quad\includegraphics[width=0.5\textwidth,clip]{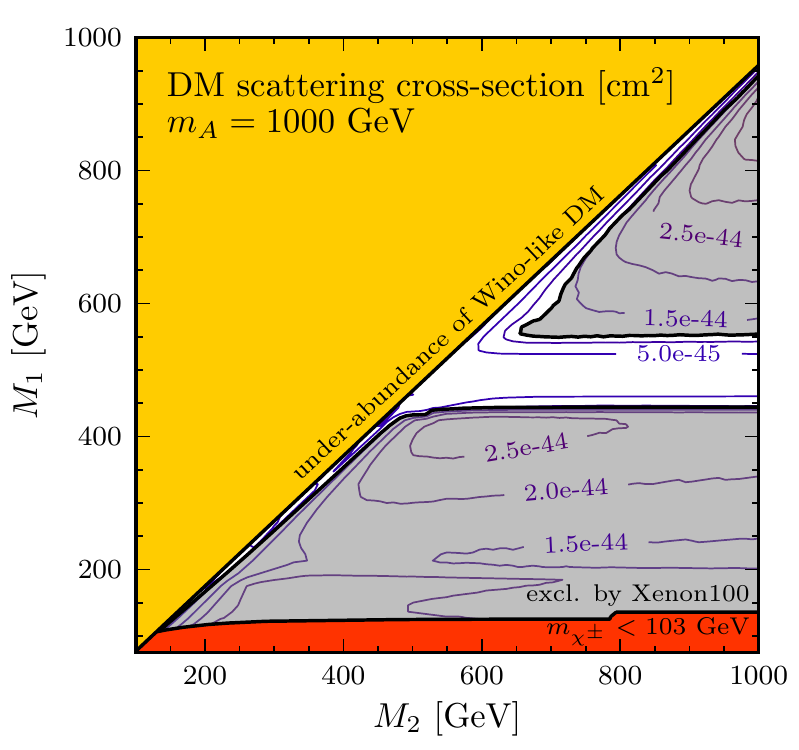}}
\caption{\label{fig:direct}\it\small Curves of constant neutralino-proton elastic spin-independent scattering cross section, for the ``well-tempered'' neutralinos of Fig.~\ref{fig:isomu}, for $m_A=500$ GeV (left) and 1000 GeV (right).}
\end{figure*}

We calculate in this section rates for the direct and indirect detection of dark matter. We start with the exploration, in Fig.\ref{fig:direct}, of the spin-independent neutralino-proton scattering cross section, for which we show several iso-level curves\footnote{For these and all other dark matter detection cross sections and rates, we employ the DarkSUSY code \cite{darksusy} with default parameters for the Galactic dark matter halo, quark content of the proton, etc.}. In the present framework, this quantity depends on the coupling of the lightest neutralino to the CP-even Higgses $h$ and $H$ and thus, in turn, on the lightest neutralino's higgsino fraction. Small higgsino fraction, as encountered near the $M_1\simeq M_2$ border and in the $m_\chi\simeq m_A/2$ resonance region, suppresses the scattering off of nucleons. The larger the higgsino mixing, the larger the cross section (which at large $\tan\beta$ is proportional to $(N_{11}(N_{12}-N_{13}))^2$, where $N$ is the matrix that diagonalizes the neutralino mass matrix), as can be appreciated by noticing the increase in the cross section with $M_1$, which corresponds to values of $\mu$ that are increasingly more degenerate with $M_1$ in order to satisfy the relic density constraint.

We shade in grey the region that is already excluded by current, recent results from the Xenon100 experiment \cite{xenon100}. The recent results from 225 live days represent a very significant improvement over the previous years' results \cite{xenon100old}, with an important impact on the regions ruled out by direct dark matter searches. The region ruled out corresponds to $M_1\lesssim220$ GeV and $M_1\gtrsim300-350$ GeV for $m_A=500$ GeV, and to $M_1\lesssim450$ GeV or $M_1\gtrsim550$ GeV for $m_A=1000$ GeV (with the exception of the narrow regions at $M_1\simeq M_2$). The pattern observed for the two values of $m_A$ under consideration here continues for other values of $m_A$, leaving strips 60-100 GeV wide around $M_1=m_A/2$. 

\begin{figure*}[!t]
\mbox{\hspace*{-1.2cm}\includegraphics[width=0.55\textwidth,clip]{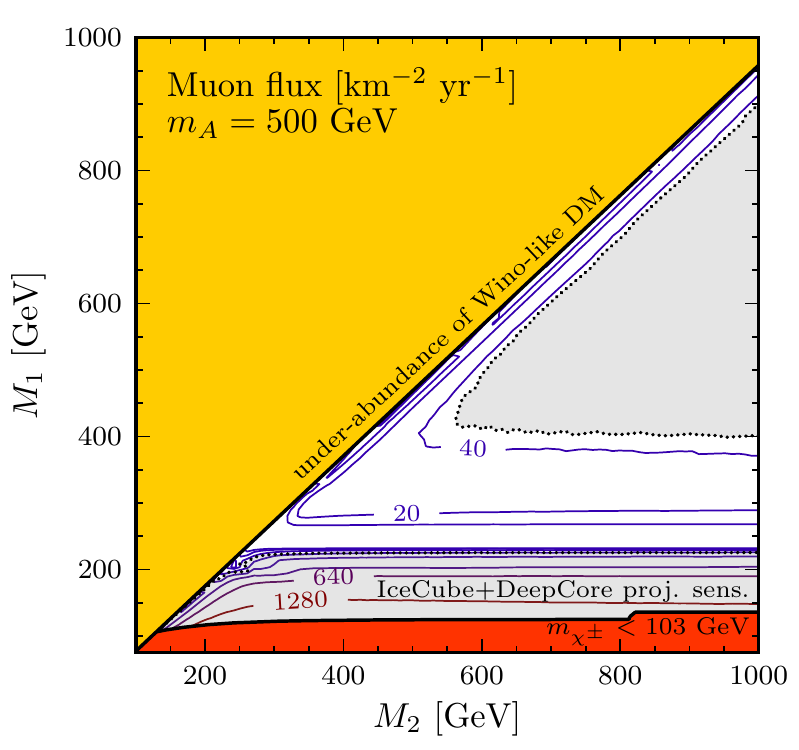}\quad\includegraphics[width=0.55\textwidth,clip]{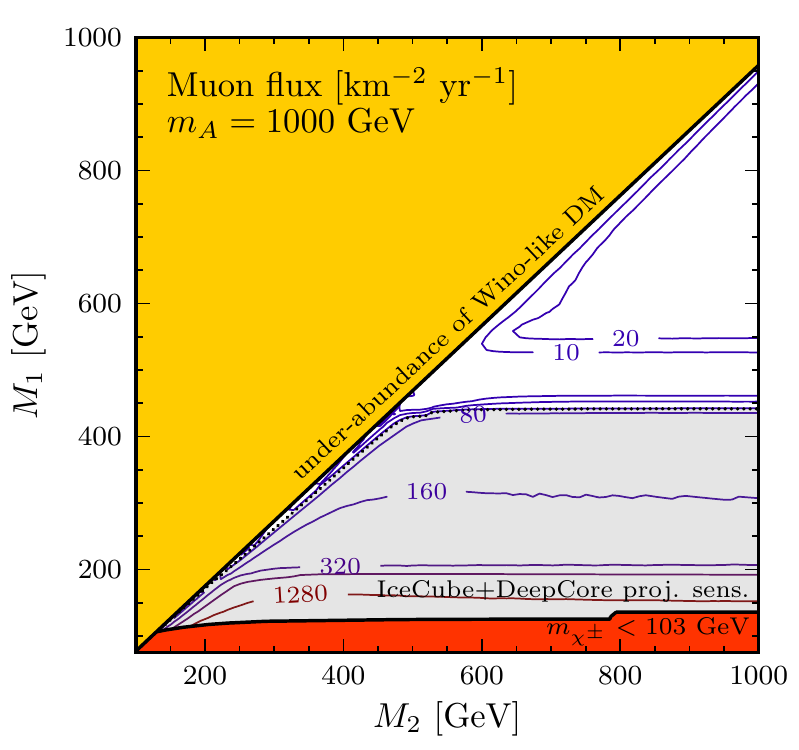}}
\caption{\label{fig:NT}\it\small Curves of constant flux of muons from neutrinos produced by neutralino pair-annihilation in the Sun, for the ``well-tempered'' neutralinos of Fig.~\ref{fig:isomu}.}
\end{figure*}

We now turn to indirect detection, and specifically to the search for high-energy neutrinos from the Sun produced by dark matter annihilation. In the very near future, the now-completed km$^3$ high-energy neutrino detector IceCube and the compact Cherenkov detector DeepCore operating at IceCube's center (and featuring a comparatively much lower energy threshold), will deliver data of great relevance in the search for particle dark matter. For the parameter space of interest here, the key search will be targeting high-energy neutrinos originating from neutralino pair-annihilation at the core of the Sun, where the neutralinos had been trapped by successive scattering with the Sun's nuclei.

Fig.~\ref{fig:NT} shows the predicted integrated muon flux for muons resulting from muon neutrino charged-current interactions, integrated above a conventional 1 GeV threshold\footnote{We note that IceCube has a much higher energy threshold, but the combined reach of IceCube-DeepCore is customarily expressed in terms of the integrated flux above 1 GeV, for a given final state (in this case pair-annihilation into $W^+W^-$ and $ZZ$).} for 180 live-days (roughly one year of operations), from neutrinos produced by neutralino pair-annihilation in the Sun. Current constraints from operating neutrino telescopes \cite{icecube} do not exclude any of the shown parameter space. We shade in grey the region that will be probed with 180 days of IceCube80 plus Deep-Core data \cite{icecube}. The neutrino flux from the Sun depends primarily on the capture rate in the Sun which, in turn, is highly sensitive to the spin-dependent neutralino-nucleon cross section. Again, this cross section depends on the higgsino fraction, and is suppressed in the regions of wino-bino mixing as well as in those where the mechanism that sets the correct neutralino thermal relic density is resonant annihilation, and where the higgsino fraction is much lower (see Fig.~\ref{fig:isomu}). Future prospects for neutrino telescopes are, overall, rather promising, covering most of the parameter space where resonant pair-annihilation does not occur. Current direct detection results (Fig.~\ref{fig:direct}), however, exclude the possibility to have a signal from neutrino telescopes in this model at the sensitivity level under consideration here.

\begin{figure*}[!t]
\mbox{\hspace*{-1.2cm}\includegraphics[width=0.55\textwidth,clip]{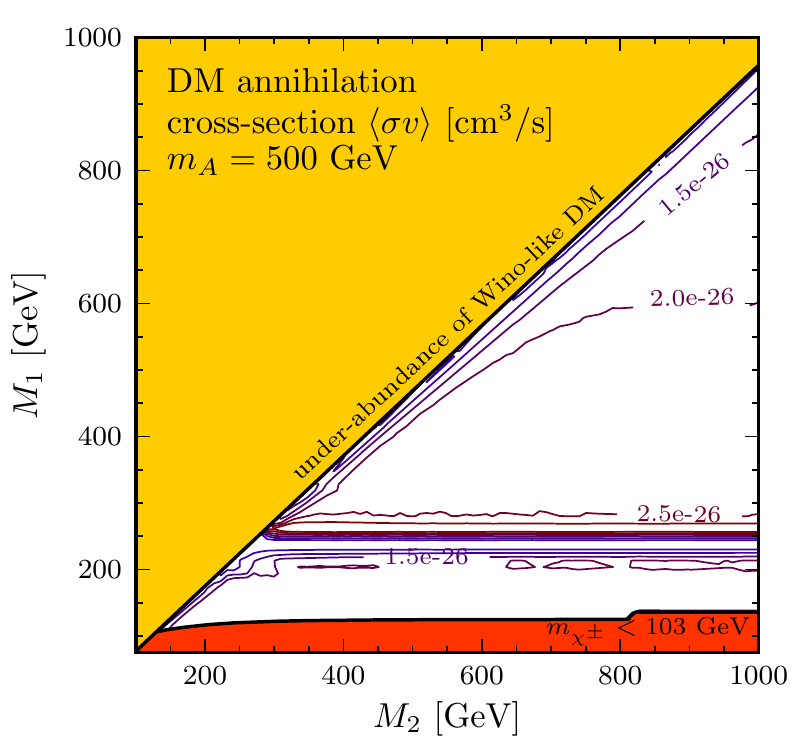}\quad\includegraphics[width=0.55\textwidth,clip]{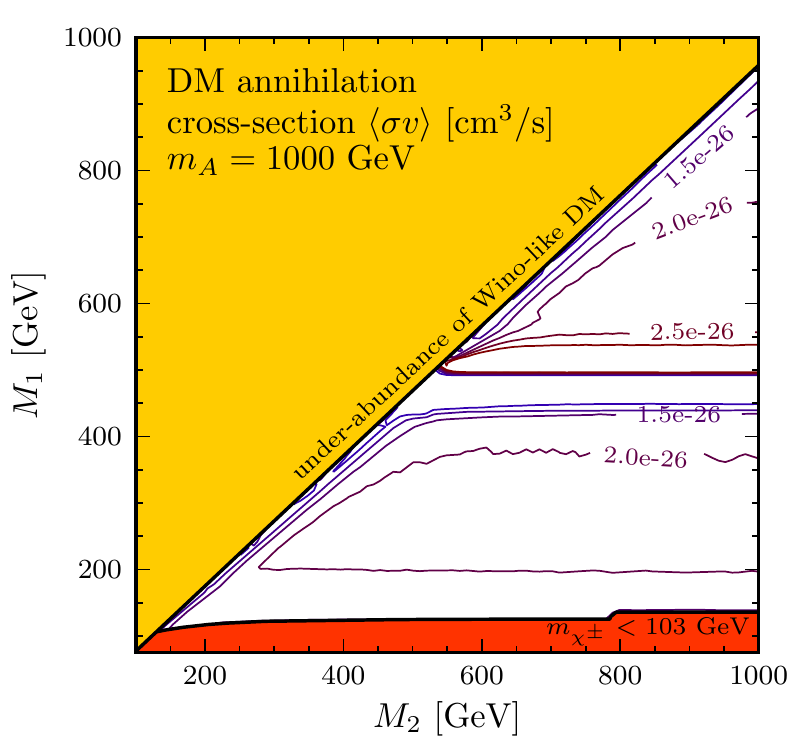}}
\caption{\label{fig:sigmav}\it\small Curves of constant neutralino zero-temperature thermally averaged pair-annihilation cross section $\langle\sigma v\rangle$, for the ``well-tempered'' neutralinos of Fig.~\ref{fig:isomu}.}
\end{figure*}

Finally, in Fig.~\ref{fig:sigmav} we concentrate on other indirect detection methods, such as the search for gamma-rays or of antimatter resulting from the pair-annihilation of dark matter in the Galactic halo. All of the associated rates depend linearly upon the (zero temperature, thermally averaged) pair-annihilation cross section $\langle\sigma v\rangle$, and, generically, on the inverse of the dark matter particle mass. We find, as expected, that the neutralino pair annihilation cross section is uniformly close to $\langle\sigma v\rangle\simeq (1-2.5)\times 10^{-26}{\rm cm}^3/{\rm s}$; we note that this value differs from the canonical $\langle\sigma v\rangle\simeq 3\times 10^{-26}{\rm cm}^3/{\rm s}$ because of neutralino co-annihilation with the chargino and next-to-lightest neutralino that contribute significantly to the freeze-out process when $M_1\simeq\mu$. Given the relatively large values of the neutralino masses in the regions of interest, $m_\chi\gtrsim 100$ GeV, all parameter space is beyond current constraints from gamma-rays \cite{dsphnew} (constraints from antimatter are highly dependent on assumptions on cosmic ray propagation, and are not tighter than those from gamma-rays for conservative choices of the parameters describing cosmic-ray Galactic diffusion). The low-$M_1$ region might be testable with increased statistics from the Fermi Large Area Telescope \cite{Ackermann:2011wa}. Going beyond the projected reach of neutrino telescopes would, however, need an improvement of more than one order of magnitude over existing limits, which appears unrealistic in the immediate future.

\section{Non-resonant sources}\label{sec:nonressrc}

Before discussing the accidental SUSY DM and EWB parameter space, we briefly address the potential effects of non-resonant CP-violating sources on EWB.  The source we consider in Eq.~\ref{eq:source} was computed in the Higgs vev-insertion approximation, in which interactions of the higgsinos, gauginos, and Higgs vevs at the EWPT bubble wall are treated perturbatively.  This framework is also used to compute the chiral relaxation rates, whose resulting resonant structure must be taken into account to obtain more realistic estimates for the BAU.  Alternatively, one can implement a resummation of the vev-insertions for the CP-violating sources by considering the interactions with the Higgs vevs as resulting in spacetime-dependent mass matrices for the supersymmetric particles.  This was carried out in Refs.~\cite{Carena:2000id, Carena:2002ss}, which showed that the resummation effectively ``smooths out'' the resonance predicted by the vev-insertion approximation, and results in new sources that are not obtained in the vev-insertion method.  These new CP-violating sources are not resonant and are not proportional to $\Delta \beta$, hence escaping the suppression for increasing $m_A$.  While the resonant sources will dominate for regions with $M_{1,2}\sim \mu$, the non-resonant contributions may become important away from these regions. 

\begin{figure*}[!t]
\mbox{\hspace*{-1.2cm}\includegraphics[width=0.55\textwidth,clip]{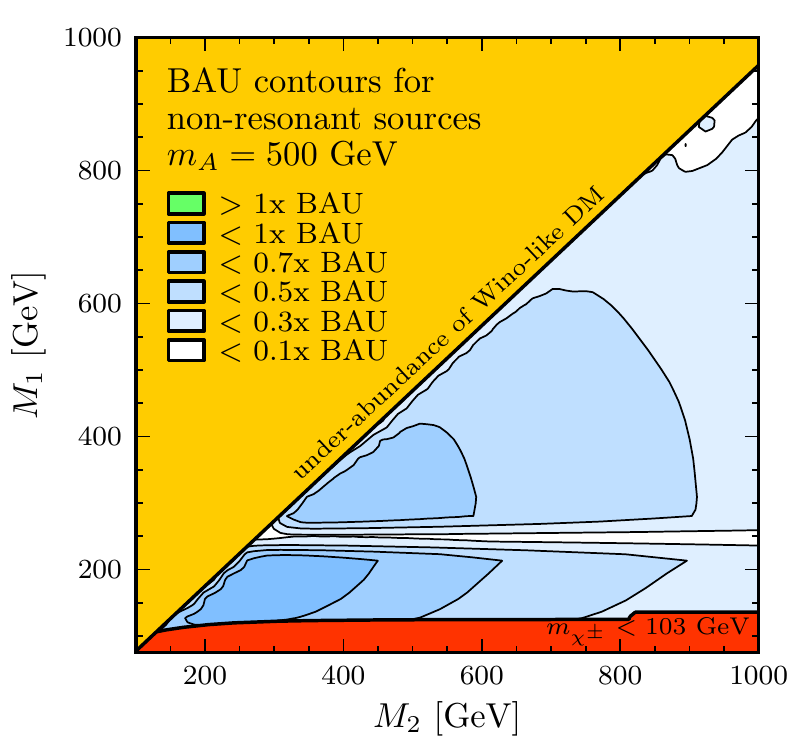}\quad\includegraphics[width=0.55\textwidth,clip]{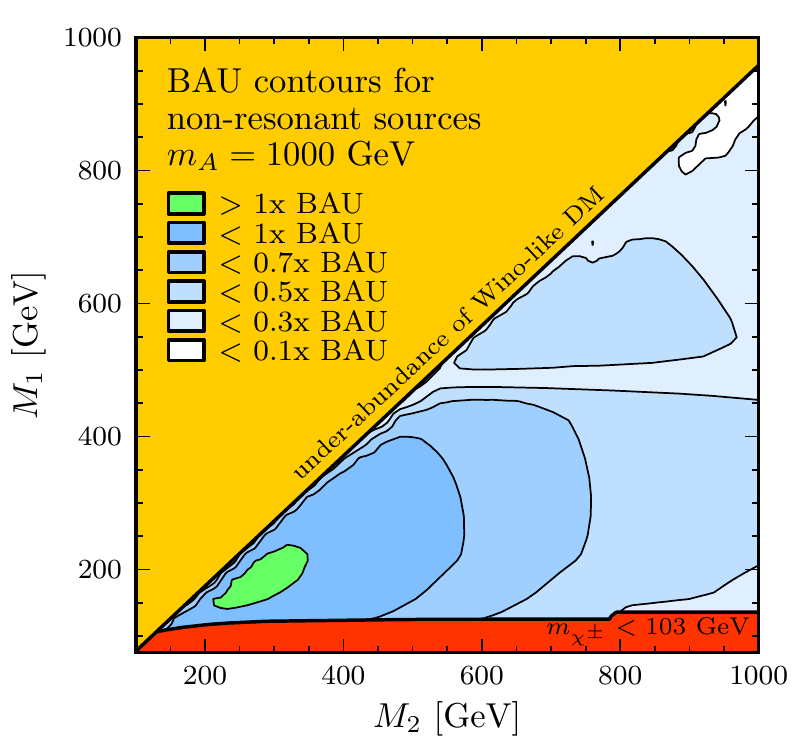}}
\caption{\label{fig:nonressrc}\it\small Curves of constant BAU from non-resonant sources, in units of the observed BAU, for the ``well-tempered'' neutralinos of Fig.~\ref{fig:isomu}, for $m_A=500$ GeV (left) and 1000 GeV (right)}
\end{figure*}

To investigate the impact of non-resonant sources on the accidental SUSY EWB parameter space, we calculate the baryon asymmetry\footnote{As in Refs.~\cite{Carena:2000id, Carena:2002ss}, we consider only the chargino contribution to the CP-violating source in this Section.} following the methods of Refs.~\cite{Carena:2000id, Carena:2002ss} with the well-tempered neutralino values in the gaugino mass planes and the same assumptions for the particle spectrum as in Sec.~\ref{sec:BAU}.  The results are shown in Fig.~\ref{fig:nonressrc} for $m_A=500, 1000$ GeV on the left and right, respectively, for maximal CP-violating phase.  As expected, the non-resonant contributions dominate away from the resonance, as can be seen by comparing Figs.~\ref{fig:EWB} and \ref{fig:nonressrc}.  However, for our choices of parameters, these sources do not open up any additional viable parameter space for EWB, other than potentially the green region on the right of Fig.~\ref{fig:nonressrc}, which is solidly ruled out by direct dark matter searches.  We discuss the effects of including non-resonant contributions on the EWB-compatible accidental SUSY parameter space below.  Note that these sources enter with {\em opposite sign} relative to the resonant sources we consider in Sec.~\ref{sec:BAU}.  

\section{The Accidental SUSY baryogenesis parameter space}\label{sec:paramspace}

\begin{figure*}[!t]
\mbox{\hspace*{-1.2cm}\includegraphics[width=0.55\textwidth,clip]{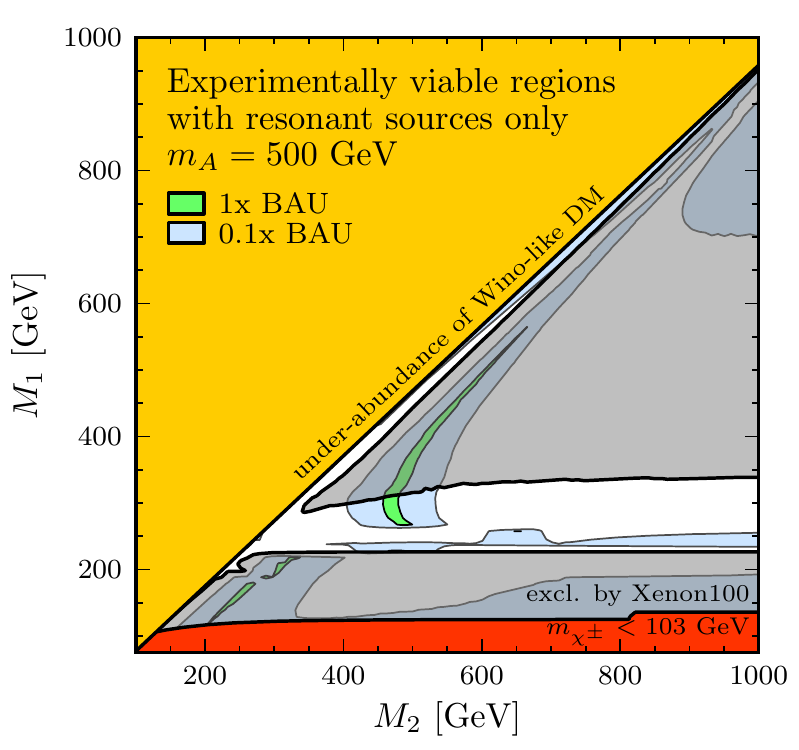}\quad\includegraphics[width=0.55\textwidth,clip]{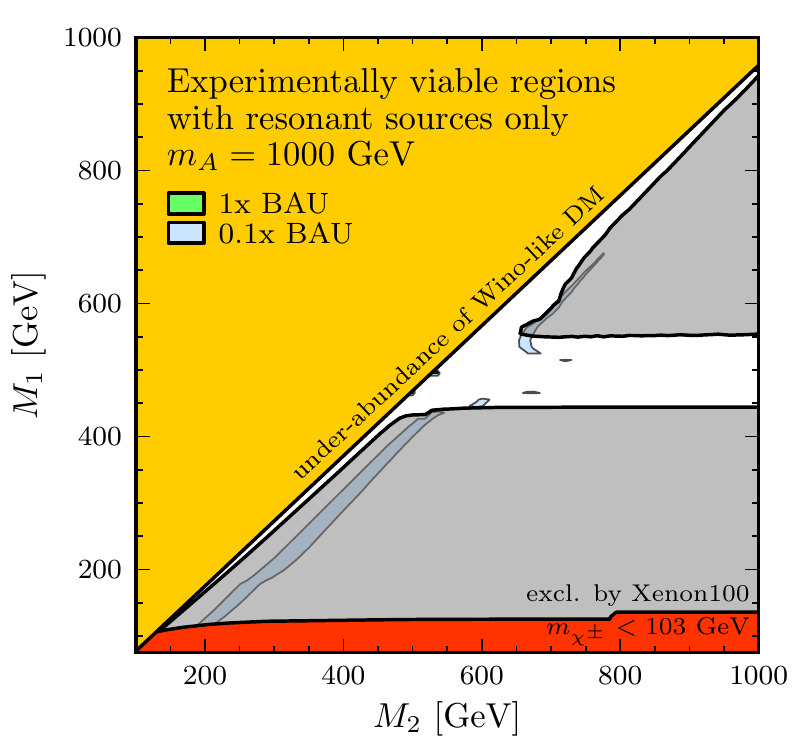}}
\caption{\label{fig:allowed}\it\small A summary plot for the parameter space of accidental supersymmetry compatible with successful electroweak baryogenesis, constraints from EDMs and dark matter searches. The green regions correspond to regions that produce 100\% of the BAU and that are compatible with EDM searches; within the light blue regions, CP-violating phases compatible with EDM constraints yield a BAU greater or equal to 10\% of the observed value.  We shade in gray the portion of parameter space ruled out by direct dark matter searches with Xenon100 \cite{xenon100}, and as in all other plots we set $m_A=500$ GeV in the left panel and 1000 GeV in the right panel.}
\end{figure*}

\begin{figure*}[!t]
\mbox{\hspace*{-1.2cm}\includegraphics[width=0.55\textwidth,clip]{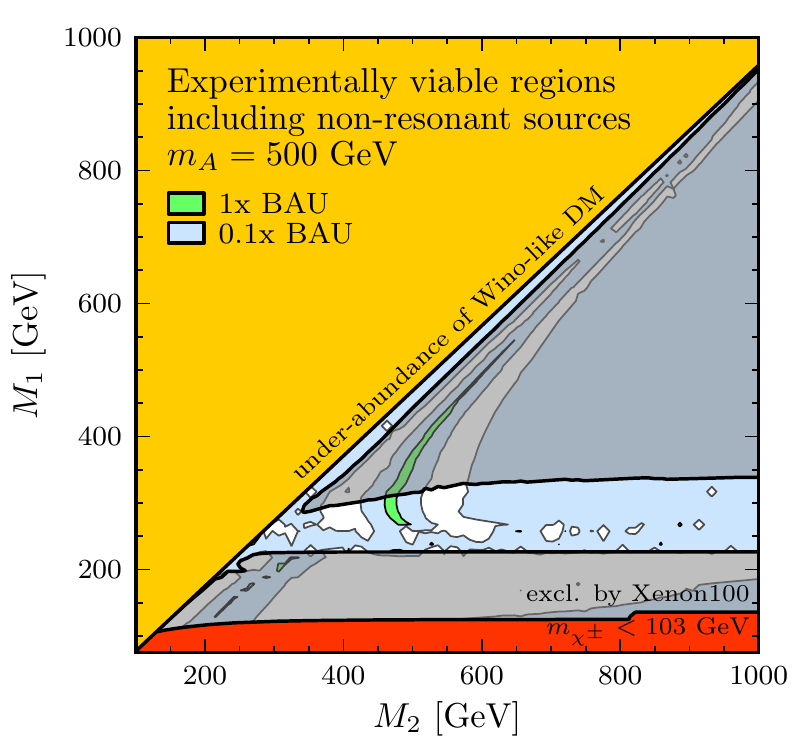}\quad\includegraphics[width=0.55\textwidth,clip]{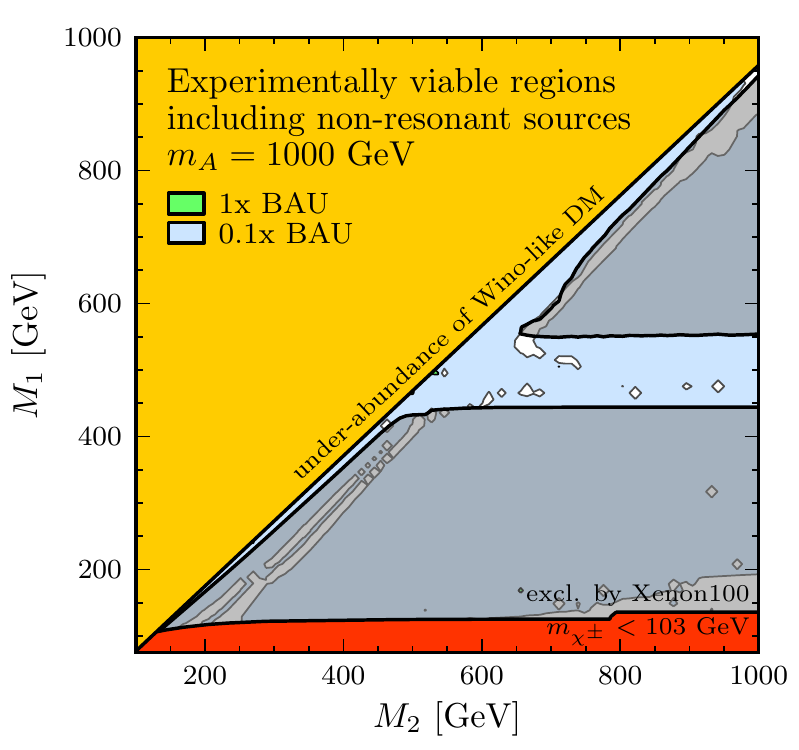}}
\caption{\label{fig:allowed_nonres}\it\small Same as in Fig.~\ref{fig:allowed}, but now including non-resonant sources, for $m_A=500$ GeV (left) and 1000 GeV (right).}
\end{figure*}

In this section we summarize our findings, and search for the portions of the parameter space of accidental supersymmetry that produce both a good thermal relic neutralino abundance and successful baryogenesis at the electroweak phase transition. To do so, we consider both resonant sources only (Fig.~\ref{fig:allowed}) and resonant plus non-resonant sources (Fig.~\ref{fig:allowed_nonres}). We calculate the maximal BAU that can be produced at each parameter space point such that constraints from EDM searches are not violated, and we superimpose limits deriving from dark matter direct searches with Xenon100. As noted above, results for intermediate values of $m_A$ interpolate between what we find for the two specific values chosen here.

The key findings of this section are that: 
\begin{itemize}
\item the lightest neutralino mass must have a mass between 200 and 500 GeV 
\item the masses of all charginos and neutralinos lie within a factor 2 of the lightest neutralino mass
\item the heavy Higgs sector must be below 1 TeV (no viable parameter space is open for $m_A\gtrsim1$ TeV, see the right panel of Fig.~\ref{fig:allowed} and \ref{fig:allowed_nonres}) and lies within approximately 20-25\% of twice the lightest neutralino mass (to comply with direct detection constraints)
\end{itemize}

Fig.~\ref{fig:allowed} shows the allowed parameter space that is consistent with the observed dark matter relic abundance, electroweak baryogenesis, EDM constraints, and dark matter direct detection constraints, ignoring the contribution of non-resonant sources. As in all other plots, at each point on the $M_1$--$M_2$ plane, $\mu$ is set to give the correct dark matter abundance. The phases $\phi_{M_1}=\phi_{M_2}$ are set to the maximal value compatible with EDM searches. The green central region is consistent with all of the baryon asymmetry coming from the electroweak phase transition, whereas in the larger blue regions electroweak baryogenesis would only account for a fraction of the asymmetry, at least 10\%, unless a correction of order unity is applied to the calculation of the BAU. 

Across the parameter space we consider here, the only viable regions are those for which the baryon asymmetry (as calculated, for a maximal CP-violating phase, in Fig.~\ref{fig:EWB})  is large enough such that the regions still satisfy the BAU requirement when the CP-phase is reduced to avoid the EDM constraints. For $m_A = 1000$ GeV, we find no viable region of parameter space satisfying all requirements we impose. We note that $M_1$ (and thus the lightest neutralino mass, in the parameter space of interest here) ranges between 200 (for smaller values of $m_A$, consistent with particle physics constraints such as e.g. those arising from $b\to s\gamma$) and 500 GeV; $\mu$ and $M_2$ are both within a factor 2 of $M_1$, with a degree of degeneracy that increases with increasing mass. As a result, all four neutralinos and two charginos in the electroweak ``-ino'' sector are compressed to within a factor 2 of the lightest neutralino mass, whose value is, in turn, constrained to $200\lesssim m_\chi/{\rm GeV}\lesssim 500$. Also, Fig.~\ref{fig:allowed} and \ref{fig:allowed_nonres} illustrate that direct detection constraints imply that the heavy Higgs sector lie within 20-25\% of twice the LSP mass.

Fig.~\ref{fig:allowed_nonres} adds to the calculation of the BAU the non-resonant sources contribution discussed in sec.~\ref{sec:nonressrc}. Non-resonant sources play an important role in opening up most of the accidental SUSY parameter space outside the regions where resonant terms (with which they negatively interfere) are important. Still, non-resonant contributions fall short of producing 100\% of the observed BAU, which is again limited to an even narrower region of parameter space with $M_1\sim M_2\sim \mu$. Also, we find again that no parameter space is open at $m_A=1$ TeV if one insists on successful EWB. Qualitatively, the addition of non-resonant sources does not change our conclusions.

\section{Discussion and Conclusions}\label{sec:conclusions}
Accidental supersymmetry is a particle physics framework that naturally addresses both the large and the little hierarchy problems as well as the potential CP and flavor problems of supersymmetry, while in principle providing a successful thermal dark matter candidate. We argued here that this framework naturally accommodates successful electroweak baryogenesis, for the following reasons:
\begin{itemize}
\item[(i)] a strongly first order electroweak phase transition may be a generic feature of this framework, either as a consequence of supercooling produced by the phase transition between the high and low-temperature RS spacetimes, or from the contribution of a singlet to the superpotential as may be required to solve the $\mu$-problem;
\item[(ii)] light third-generation stops and gauginos allow for resonant CP-violating sources to produce potentially large net chiral currents fueling a large enough net baryon number via sphaleron transitions; 
\item[(iii)] heavy first- and second-generation sfermions prevent excessive one-loop contributions to observable electric dipole moments in the presence of the needed large CP-violating phases.
\end{itemize}
Here, we carried out a model-independent study (from the standpoint of supersymmetry breaking), although for definiteness we picked a specific accidental SUSY spectrum realization. Specifically, we let the relevant U(1)$_Y$ and SU(2) gaugino soft supersymmetry breaking masses $M_1$ and $M_2$, as well as the higgsino mass parameter $\mu$ vary freely. We constrained this triplet of mass parameters enforcing that the lightest supersymmetric particle be a neutralino with a thermal relic density matching the observed density of dark matter. In practice, this amounted to selecting values of $\mu$ across the $(M_1,\ M_2)$ parameter space so that the higgsino fraction drove the thermal relic density of the lightest neutralino to the desired value.

After enforcing the relic density constraint, we proceeded to calculate the baryon asymmetry resulting from electroweak baryogenesis across the $(M_1,\ M_2)$ parameter space. We included both resonant and non-resonant sources, and we picked two representative values for the heavy Higgs sector mass scale, which is relevant for resonant sources. The requirement of successful baryogenesis generically restricted the viable parameter space to a relatively narrow funnel at $M_1\lesssim M_2$, with $\mu\sim M_1,\ M_2$; recent direct detection constraints also enforce $M_1\simeq m_A/2$ to within 20-25\%.

The strongest constraints on this framework derive from the non-observation of electric dipole moments and of signals from dark matter direct detection, most notably with the Xenon100 experiment \cite{xenon100}. We calculated in detail how these constraints restrict the parameter space relevant for baryogenesis, concluding that dark matter direct searches eliminate neutralinos with a large higgsino fraction (requiring to some degree resonant annihilation through the heavy Higgs sector, and hence $M_1\simeq m_A/2$), while electric dipole moments greatly restrict regions of viable electroweak baryogenesis to those parameter space points producing, for maximal CP violating phases, a BAU much larger than observed (those parameter space regions are then compatible with successful baryogenesis as the CP phases are lowered to comply with EDM searches).

We calculated the predicted EDM and dark matter search rates in the framework of accidental supersymmetric baryogenesis and we concluded that:
\begin{itemize}
\item the most sensitive EDM search to constrain this model is provided by searches for the electron EDM; an improvement of one order of magnitude on the current experimental sensitivity would conclusively test the framework, even allowing for some theoretical uncertainties in the calculation of the BAU;
\item the entire parameter is highly constrained by current direct, spin-independent dark matter-nucleon cross section limits, and will soon be fully tested even for resonant neutralino annihilation
\item the predicted signal at neutrino telescopes from neutralino annihilation in the core of the Sun is potentially large enough for detection, although direct detection results imply that no signal is expected within approximately one year of data taking
\end{itemize}

The parameter space compatible with successful baryogenesis and thermal dark matter is highly constrained, and is characterized by a lightest neutralino with a mass between 200 and 500 GeV, with all other neutralino and chargino masses within a factor 2 of the lightest neutralino mass, with $m_A\simeq 2M_1<1$ TeV. This compressed electroweak ``inos'' spectrum might be challenging for LHC searches, but would be ideally targeted with an $e^+ e^-$ linear collider with a TeV center of mass energy.

Concluding, we demonstrated here that accidental supersymmetry is an explicit realization of a framework for successful thermal relic dark matter and electroweak baryogenesis, which is motivated by an entirely different set of theoretical arguments based upon addressing the hierarchy, CP and flavor problems. We showed that accidental supersymmetric baryogenesis is a highly constrained setup, but one with very sharp experimental predictions for electric dipole moment, dark matter, and collider searches. We therefore anticipate that this scenario be falsified or produce signals in the very near future in a variety of experiments.

\begin{acknowledgments}
\noindent  We thank Tony Gherghetta and Michael J. Ramsey-Musolf for helpful discussions and feedback.  SP and JK are partly supported by an Outstanding Junior Investigator Award from the US Department of Energy and by Contract DE-FG02-04ER41268, and by NSF Grant PHY-0757911. CLW is supported by an NSF graduate fellowship.

\end{acknowledgments}


\begin{thebibliography}{300}

\bibitem{Randall:1999ee} 
  L.~Randall and R.~Sundrum,
  Phys.\ Rev.\ Lett.\  {\bf 83}, 3370 (1999)
  [hep-ph/9905221].
  
\bibitem{Gherghetta:2000qt} 
  T.~Gherghetta and A.~Pomarol,
  Nucl.\ Phys.\ B {\bf 586}, 141 (2000)
  [hep-ph/0003129].
  
  \bibitem{KK_modes} 
  K.~Agashe, G.~Perez and A.~Soni,
  Phys.\ Rev.\ D {\bf 71}, 016002 (2005)
  [hep-ph/0408134].
  
  
  \bibitem{Gherghetta:2003wm} 
  T.~Gherghetta and A.~Pomarol,
  Phys.\ Rev.\ D {\bf 67}, 085018 (2003)
  [hep-ph/0302001].
  
  \bibitem{Sundrum:2009gv} 
  R.~Sundrum,
  JHEP {\bf 1101}, 062 (2011)
  [arXiv:0909.5430 [hep-th]].
  
\bibitem{accidental} 
  T.~Gherghetta, B.~von Harling and N.~Setzer,
  JHEP {\bf 1107}, 011 (2011)
  [arXiv:1104.3171 [hep-ph]].
  
  \bibitem{Sakharov:1967dj} 
  A.~D.~Sakharov,
  Pisma Zh.\ Eksp.\ Teor.\ Fiz.\  {\bf 5}, 32 (1967)
  [JETP Lett.\  {\bf 5}, 24 (1967)]
  [Sov.\ Phys.\ Usp.\  {\bf 34}, 392 (1991)]
  [Usp.\ Fiz.\ Nauk {\bf 161}, 61 (1991)].
  
  \bibitem{Nardini:2007me} 
  G.~Nardini, M.~Quiros and A.~Wulzer,
  JHEP {\bf 0709}, 077 (2007)
  [arXiv:0706.3388 [hep-ph]].
  
  \bibitem{Creminelli:2001th} 
  P.~Creminelli, A.~Nicolis and R.~Rattazzi,
  JHEP {\bf 0203}, 051 (2002)
  [hep-th/0107141].
  
 \bibitem{Randall:2006py} 
  L.~Randall and G.~Servant,
  JHEP {\bf 0705}, 054 (2007)
  [hep-ph/0607158].
  
  \bibitem{Kaplan:2006yi} 
  J.~Kaplan, P.~C.~Schuster and N.~Toro,
  hep-ph/0609012.
  
\bibitem{Goldberger:1999uk} 
  W.~D.~Goldberger and M.~B.~Wise,
  Phys.\ Rev.\ Lett.\  {\bf 83}, 4922 (1999)
  [hep-ph/9907447].
 
   \bibitem{Carena:2008vj}
  M.~Carena, G.~Nardini, M.~Quiros, C.~E.~M.~Wagner,
  Nucl.\ Phys.\  {\bf B812}, 243-263 (2009).
  [arXiv:0809.3760 [hep-ph]].
  
  \bibitem{Carena:2008rt} 
  M.~Carena, G.~Nardini, M.~Quiros and C.~E.~M.~Wagner,
  JHEP {\bf 0810}, 062 (2008)
  [arXiv:0806.4297 [hep-ph]].
  
  \bibitem{Profumo:2007wc}
  S.~Profumo, M.~J.~Ramsey-Musolf, G.~Shaughnessy,
  JHEP {\bf 0708}, 010 (2007).
  [arXiv:0705.2425 [hep-ph]].

  \bibitem{Kozaczuk:2012xv} 
  J.~Kozaczuk, S.~Profumo, M.~J.~Ramsey-Musolf and C.~L.~Wainwright,
  arXiv:1206.4100 [hep-ph].

 \bibitem{Huet:1995sh}
  P.~Huet, A.~E.~Nelson,
  Phys.\ Rev.\  {\bf D53}, 4578-4597 (1996).
  [hep-ph/9506477].

\bibitem{Carena:1996wj} 
 M.~S.~Carena, M.~Quiros and C.~E.~M.~Wagner,
  Phys.\ Lett.\ B {\bf 380}, 81 (1996)
  [hep-ph/9603420].


   \bibitem{Lee:2004we}
  C.~Lee, V.~Cirigliano and M.~J.~Ramsey-Musolf,
  Phys.\ Rev.\  D {\bf 71}, 075010 (2005)
  [arXiv:hep-ph/0412354].
  
  \bibitem{Chung:2008aya}
  D.~J.~H.~Chung, B.~Garbrecht, M.~J.~Ramsey-Musolf, S.~Tulin,
  Phys.\ Rev.\ Lett.\  {\bf 102}, 061301 (2009).
  [arXiv:0808.1144 [hep-ph]].
  
   \bibitem{Lepton_Mediated}
  D.~J.~H.~Chung, B.~Garbrecht, M.~J.~Ramsey-Musolf and S.~Tulin,
  Phys.\ Rev.\  D {\bf 81}, 063506 (2010)
  [arXiv:0905.4509 [hep-ph]].
  
  \bibitem{Supergauge}
  D.~J.~H.~Chung, B.~Garbrecht, M.~J.~Ramsey-Musolf and S.~Tulin,
  Phys.\ Rev.\ Lett.\  {\bf 102}, 061301 (2009)
  [arXiv:0808.1144 [hep-ph]].

  \bibitem{Including_Yukawa}
  V.~Cirigliano, M.~J.~Ramsey-Musolf, S.~Tulin and C.~Lee,
  Phys.\ Rev.\  D {\bf 73}, 115009 (2006)
  [arXiv:hep-ph/0603058].

  \bibitem{Konstandin:2003dx}
  T.~Konstandin, T.~Prokopec and M.~G.~Schmidt,
  Nucl.\ Phys.\  B {\bf 679}, 246 (2004)
  [arXiv:hep-ph/0309291].

  \bibitem{Konstandin:2004gy}
  T.~Konstandin, T.~Prokopec, M.~G.~Schmidt,
  Nucl.\ Phys.\  {\bf B716}, 373-400 (2005).
  [hep-ph/0410135].
  
   \bibitem{More_Relaxed}
  A.~Riotto,
  Phys.\ Rev.\  D {\bf 58}, 095009 (1998)
  [arXiv:hep-ph/9803357].
  
  \bibitem{Carena:2002ss}
  M.~S.~Carena, M.~Quiros, M.~Seco, C.~E.~M.~Wagner,
  Nucl.\ Phys.\  {\bf B650}, 24-42 (2003).
  [hep-ph/0208043].
  
\bibitem{EWB_and_EDMs}
  V.~Cirigliano, Y.~Li, S.~Profumo and M.~J.~Ramsey-Musolf,
  JHEP {\bf 1001}, 002 (2010)
  [arXiv:0910.4589 [hep-ph]].

    \bibitem{Balazs:2004ae}
  C.~Balazs, M.~S.~Carena, A.~Menon, D.~E.~Morrissey, C.~E.~M.~Wagner,
  Phys.\ Rev.\  {\bf D71}, 075002 (2005).
  [hep-ph/0412264].
  
  \bibitem{EWB_and_DM}
  V.~Cirigliano, S.~Profumo and M.~J.~Ramsey-Musolf,
  JHEP {\bf 0607}, 002 (2006)
  [arXiv:hep-ph/0603246].
  
  \bibitem{Kozaczuk:2011vr} 
  J.~Kozaczuk and S.~Profumo,
  JCAP {\bf 1111}, 031 (2011)
  [arXiv:1108.0393 [hep-ph]].
  
  \bibitem{Menon:2004wv} 
  A.~Menon, D.~E.~Morrissey and C.~E.~M.~Wagner,
  Phys.\ Rev.\ D {\bf 70}, 035005 (2004)
  [hep-ph/0404184].
  
\bibitem{Huber:2006wf} 
  S.~J.~Huber, T.~Konstandin, T.~Prokopec and M.~G.~Schmidt,
  Nucl.\ Phys.\ B {\bf 757}, 172 (2006)
  [hep-ph/0606298].
  
  \bibitem{welltempered} 
  N.~Arkani-Hamed, A.~Delgado and G.~F.~Giudice,
  Nucl.\ Phys.\ B {\bf 741}, 108 (2006)
  [hep-ph/0601041].
   

  \bibitem{split_susy}
  N.~Arkani-Hamed and S.~Dimopoulos,
  JHEP {\bf 0506}, 073 (2005)
  [hep-th/0405159].;
   G.~F.~Giudice and A.~Romanino,
  Nucl.\ Phys.\ B {\bf 699}, 65 (2004)
  [Erratum-ibid.\ B {\bf 706}, 65 (2005)]
  [hep-ph/0406088].
  N.~Arkani-Hamed, S.~Dimopoulos, G.~F.~Giudice and A.~Romanino,
  Nucl.\ Phys.\ B {\bf 709}, 3 (2005)
  [hep-ph/0409232].

\bibitem{Casas:2001xv} 
  J.~A.~Casas, J.~R.~Espinosa and I.~Navarro,
  Nucl.\ Phys.\ B {\bf 620}, 195 (2002)
  [hep-ph/0109127].
  
  \bibitem{Goh:2003yr} 
  H.~-S.~Goh, M.~A.~Luty and S.~-P.~Ng,
  JHEP {\bf 0501}, 040 (2005)
  [hep-th/0309103].
  
  \bibitem{radion_collider} 
  M.~Chaichian, A.~Datta, K.~Huitu and Z.~-h.~Yu,
  Phys.\ Lett.\ B {\bf 524}, 161 (2002)
  [hep-ph/0110035];
  V.~Barger, M.~Ishida and W.~-Y.~Keung,
  Phys.\ Rev.\ Lett.\  {\bf 108}, 101802 (2012)
  [arXiv:1111.4473 [hep-ph]];
  H.~de Sandes and R.~Rosenfeld,
  Phys.\ Rev.\ D {\bf 85}, 053003 (2012)
  [arXiv:1111.2006 [hep-ph]];
  B.~Grzadkowski, J.~F.~Gunion and M.~Toharia,
  Phys.\ Lett.\ B {\bf 712}, 70 (2012)
  [arXiv:1202.5017 [hep-ph]].
  
  \bibitem{higgsclaims}
   [ATLAS Collaboration],
  arXiv:1202.1408 [hep-ex];
  [ATLAS Collaboration],
  arXiv:1202.1414 [hep-ex];
 [ATLAS Collaboration],
  arXiv:1202.1415 [hep-ex];
   S.~Chatrchyan {\it et al.}  [CMS Collaboration],
  arXiv:1202.1487 [hep-ex];
    S.~Chatrchyan {\it et al.}  [CMS Collaboration],
  arXiv:1202.1416 [hep-ex];
   S.~Chatrchyan {\it et al.}  [CMS Collaboration],
  arXiv:1202.1488 [hep-ex].
 
  \bibitem{Djouadi:2005gj} 
  A.~Djouadi,
  Phys.\ Rept.\  {\bf 459}, 1 (2008)
  [hep-ph/0503173].
 

\bibitem{ref22}
 J.~L.~Feng, K.~T.~Matchev and F.~Wilczek,
  Phys.\ Lett.\ B {\bf 482}, 388 (2000)
  [hep-ph/0004043].

\bibitem{ref23}
 H.~Baer, A.~Mustafayev, S.~Profumo, A.~Belyaev and X.~Tata,
  JHEP {\bf 0507}, 065 (2005)
  [hep-ph/0504001].

\bibitem{ref16}
G.~F.~Giudice and A.~Romanino,
  Nucl.\ Phys.\ B {\bf 699}, 65 (2004)
  [Erratum-ibid.\ B {\bf 706}, 65 (2005)]
  [hep-ph/0406088].

\bibitem{ref24}
 A.~Pierce,
  Phys.\ Rev.\ D {\bf 70}, 075006 (2004)
  [hep-ph/0406144].
  
\bibitem{ref25}
 A.~Masiero, S.~Profumo and P.~Ullio,
  Nucl.\ Phys.\ B {\bf 712}, 86 (2005)
  [hep-ph/0412058].
  
\bibitem{ref26}
 H.~Baer, A.~Mustafayev, E.~-K.~Park and S.~Profumo,
  JHEP {\bf 0507}, 046 (2005)
  [hep-ph/0505227];
  H.~Baer, T.~Krupovnickas, A.~Mustafayev, E.~-K.~Park, S.~Profumo and X.~Tata,
  JHEP {\bf 0512}, 011 (2005)
  [hep-ph/0511034].
  
\bibitem{ref27}
 A.~Birkedal-Hansen and B.~D.~Nelson,
  Phys.\ Rev.\ D {\bf 64}, 015008 (2001)
  [hep-ph/0102075];
  A.~Birkedal-Hansen and B.~D.~Nelson,
  Phys.\ Rev.\ D {\bf 67} (2003) 095006
  [hep-ph/0211071].
  
\bibitem{darksusy}
 P.~Gondolo, J.~Edsjo, P.~Ullio, L.~Bergstrom, M.~Schelke and E.~A.~Baltz,
  JCAP {\bf 0407}, 008 (2004)
  [astro-ph/0406204].
   
  \bibitem{Morrissey:2012db} 
  D.~E.~Morrissey and M.~J.~Ramsey-Musolf,
  arXiv:1206.2942 [hep-ph].
  
  \bibitem{Giudice:1993bb} 
  G.~F.~Giudice and M.~E.~Shaposhnikov,
  Phys.\ Lett.\ B {\bf 326}, 118 (1994)
  [hep-ph/9311367].
  
  \bibitem{Moore:1997im} 
  G.~D.~Moore,
  Phys.\ Lett.\ B {\bf 412}, 359 (1997)
  [hep-ph/9705248].
  
  \bibitem{Goldberger:1999un} 
  W.~D.~Goldberger and M.~B.~Wise,
  Phys.\ Lett.\ B {\bf 475}, 275 (2000)
  [hep-ph/9911457].
  
  \bibitem{Csaki:1999mp} 
  C.~Csaki, M.~Graesser, L.~Randall and J.~Terning,
  Phys.\ Rev.\ D {\bf 62}, 045015 (2000)
  [hep-ph/9911406].
  
  \bibitem{Carena:1997gx} 
  M.~S.~Carena, M.~Quiros, A.~Riotto, I.~Vilja and C.~E.~M.~Wagner,
  Nucl.\ Phys.\ B {\bf 503}, 387 (1997)
  [hep-ph/9702409].
  
  \bibitem{Carena:2000id} 
  M.~S.~Carena, J.~M.~Moreno, M.~Quiros, M.~Seco and C.~E.~M.~Wagner,
  Nucl.\ Phys.\ B {\bf 599}, 158 (2001)
  [hep-ph/0011055].
  
  \bibitem{Moreno:1998bq} 
  J.~M.~Moreno, M.~Quiros and M.~Seco,
  Nucl.\ Phys.\ B {\bf 526}, 489 (1998)
  [hep-ph/9801272].

  \bibitem{Huber:2001xf} 
  S.~J.~Huber, P.~John and M.~G.~Schmidt,
  Eur.\ Phys.\ J.\ C {\bf 20}, 695 (2001)
  [hep-ph/0101249].
  
  \bibitem{Li:2010ax} 
  Y.~Li, S.~Profumo and M.~Ramsey-Musolf,
  JHEP {\bf 1008}, 062 (2010)
  [arXiv:1006.1440 [hep-ph]].
  
\bibitem{Barr:1990vd} 
  S.~M.~Barr and A.~Zee,
  Phys.\ Rev.\ Lett.\  {\bf 65}, 21 (1990)
  [Erratum-ibid.\  {\bf 65}, 2920 (1990)].
  
\bibitem{Hudson:2011zz} 
  J.~J.~Hudson, D.~M.~Kara, I.~J.~Smallman, B.~E.~Sauer, M.~R.~Tarbutt and E.~A.~Hinds,
  Nature {\bf 473}, 493 (2011).
 
\bibitem{Baker:2006ts} 
  C.~A.~Baker, D.~D.~Doyle, P.~Geltenbort, K.~Green, M.~G.~D.~van der Grinten, P.~G.~Harris, P.~Iaydjiev and S.~N.~Ivanov {\it et al.},
  Phys.\ Rev.\ Lett.\  {\bf 97}, 131801 (2006)
  [hep-ex/0602020].

  \bibitem{Hewett:2012ns} 
  J.~L.~Hewett, H.~Weerts, R.~Brock, J.~N.~Butler, B.~C.~K.~Casey, J.~Collar, A.~de Govea and R.~Essig {\it et al.},
  arXiv:1205.2671 [hep-ex].
  
  \bibitem{Li:2008kz} 
  Y.~Li, S.~Profumo and M.~Ramsey-Musolf,
  Phys.\ Rev.\ D {\bf 78}, 075009 (2008)
  [arXiv:0806.2693 [hep-ph]].

  \bibitem{FeynHiggs}
   M.~Frank, T.~Hahn, S.~Heinemeyer, W.~Hollik, H.~Rzehak and G.~Weiglein,
  JHEP {\bf 0702}, 047 (2007)
  [hep-ph/0611326].
  G.~Degrassi, S.~Heinemeyer, W.~Hollik, P.~Slavich and G.~Weiglein,
  Eur.\ Phys.\ J.\ C {\bf 28}, 133 (2003)
  [hep-ph/0212020].
  S.~Heinemeyer, W.~Hollik and G.~Weiglein,
  Eur.\ Phys.\ J.\ C {\bf 9}, 343 (1999)
  [hep-ph/9812472].
  S.~Heinemeyer, W.~Hollik and G.~Weiglein,
  Comput.\ Phys.\ Commun.\  {\bf 124}, 76 (2000)
  [hep-ph/9812320].
  
  \bibitem{xenon100}
   E.~April {\it et al.}  [XENON100 Collaboration],
  arXiv:1207.5988 [astro-ph.CO].
  
   \bibitem{xenon100old}
  E.~Aprile {\it et al.} [ XENON100 Collaboration ],
  [arXiv:1104.2549 [astro-ph.CO]].

\bibitem{icecube}
 A.~M.~Brown and o.~b.~o.~Collaboration,
  arXiv:1012.1633 [astro-ph.HE];
 R.~Abbasi {\it et al.} [ ICECUBE Collaboration ],
  Phys.\ Rev.\ Lett.\  {\bf 102}, 201302 (2009).
  [arXiv:0902.2460 [astro-ph.CO]].

\bibitem{dsphnew}
M.~Ackermann {\it et al.}  [Fermi-LAT Collaboration],
  Phys.\ Rev.\ Lett.\  {\bf 107}, 241302 (2011)
  [arXiv:1108.3546 [astro-ph.HE]].
   
 \bibitem{Ackermann:2011wa} 
  M.~Ackermann {\it et al.}  [Fermi-LAT Collaboration],
  Phys.\ Rev.\ Lett.\  {\bf 107}, 241302 (2011)
  [arXiv:1108.3546 [astro-ph.HE]].
  
  \end{thebibliography}
\end{document}